\documentclass[%
aip,%
amsmath,amssymb,floatfix,
reprint,%
onecolumn,
jcp,%
longbibliography,
]{revtex4-1}

\usepackage{graphicx}% Include figure files
\usepackage{dcolumn}% Align table columns on decimal point
\usepackage{bm} % bold math
\usepackage{color}

\usepackage[english]{babel} 

\usepackage{amsmath,amsthm,latexsym,amssymb,amsfonts,epsfig}
\usepackage{mathtools}
\usepackage{mathrsfs} 

\usepackage[outer=1.9cm, inner=1.9cm, top=2.1cm, bottom=2.1cm]{geometry}

\newcommand{\sz}{s}
\newcommand{\cz}{c}

\usepackage{xcolor}
\definecolor{OliveGreen}{rgb}{0,0.6,0}

\newcommand{\vect}[1]{\boldsymbol{#1}}

\newcommand{\bNabla}{\boldsymbol{\nabla}}

\newcommand{\R}{\vect{r}}
\newcommand{\K}{\vect{k}}

\newcommand{\Intd}{\mathrm{d}}

\newcommand{\tv}{\tilde{v}}
\newcommand{\tP}{\tilde{P}}

\newcommand{\G}{\mathcal{G}}
\newcommand{\tG}{\tilde{\G}}

\newcommand{\sgn}{\mathrm{sgn}}

\newcommand{\arctanh}{\operatorname{arctanh}}

\newcommand{\Faxen}{Fax\'{e}n}

\def\XXint#1#2#3{{\setbox0=\hbox{$#1{#2#3}{\int}$}
     \vcenter{\hbox{$#2#3$}}\kern-.5\wd0}}

\newcommand{\bigO}{\mathcal{O}}

\newcommand{\etaEff}{\eta^\mathrm{eff}}

\usepackage[colorlinks=true,citecolor=OliveGreen,linkcolor=blue]{hyperref}

\usepackage{epstopdf}

\usepackage{csquotes}

\usepackage{type1ec} %

\allowdisplaybreaks

\begin{document}

%\preprint{APS/123-QED}

\title{Dynamics of a simple model microswimmer in an anisotropic fluid:
implications for alignment behavior and active transport in a nematic liquid crystal}

\author{Abdallah Daddi-Moussa-Ider}
\email{ider@thphy.uni-duesseldorf.de}

\affiliation
{Institut f\"{u}r Theoretische Physik II: Weiche Materie, Heinrich-Heine-Universit\"{a}t D\"{u}sseldorf, Universit\"{a}tsstra\ss e 1, 40225 D\"{u}sseldorf, Germany}

\author{Andreas M. Menzel}
\email{menzel@thphy.uni-duesseldorf.de}

\affiliation
{Institut f\"{u}r Theoretische Physik II: Weiche Materie, Heinrich-Heine-Universit\"{a}t D\"{u}sseldorf, Universit\"{a}tsstra\ss e 1, 40225 D\"{u}sseldorf, Germany}

\date{\today}

\begin{abstract}
    Several recent experiments investigate the orientational and transport behavior of self-driven bacteria and colloidal particles in nematic liquid crystals. Correspondingly, we study theoretically the dynamics of a minimal model microswimmer in a uniaxially anisotropic fluid. 
    As a first step, the hydrodynamic Green's function providing the resulting fluid flow in response to a localized force acting on the anisotropic fluid is derived analytically. 
    On this basis, the behavior of both puller- and pusher-type microswimmers in the anisotropic fluid is analyzed.
    Depending on the propulsion mechanism and the relative magnitude of different involved viscosities, we find alignment of the swimmers parallel or perpendicular to the anisotropy axis.
    Particularly, also an oblique alignment is identified under certain circumstances.
    The observed swimmer reorientation results from the hydrodynamic coupling between the self-induced fluid flow and the anisotropy of the surrounding fluid, which distorts the self-generated flow field. 
    We support parts of our results by a simplified linear stability analysis.
    Our theoretical predictions are in qualitative agreement with recent experimental observations on swimming bacteria in nematic liquid crystals.
    They support the objective of utilizing the, possibly switchable, anisotropy of a host fluid to guide individual microswimmers and active particles along a requested path, enabling controlled active transport.
\end{abstract}

\maketitle

% \pacs{}

\section{Introduction}

Active particles have the ability to move autonomously in a surrounding fluid by converting energy into directed motion.
Artificial self-propelled nano- and microscale machines hold great promise for future medical research to reach otherwise inaccessible areas of the body to perform delicate and precise tasks.
Prospective biomedical applications are precision nanosurgery, biopsy, and transport of radioactive substances to tumor areas and inflammation sites~\cite{wang12, wang13, paxton04}.
Over the last few decades, significant research efforts have been devoted to investigate the behavior of self-propelling active particles due to their importance and relevance as model systems for transport and locomotion in the micro- and nano-scale world; for recent reviews see Refs.~\onlinecite{lauga09, ramaswamy10, marchetti13, elgeti15, menzel15, bechinger16, zottl16, lauga2016ARFM, illien17}.
Unusual macroscopic signatures and intriguing spatiotemporal patterns emerge from the interaction between several active particles.
For instance, the onset of collective motion~\cite{gregoire04, chate08, baskaran09, mishra10, menzel12, heidenreich11, saintillan18}, formation of dynamic clusters~\cite{theurkauff12, svenvsek13, pleiner13, pohl14, goff16, sharifi16, scholz18}, wave patterns~\cite{liebchen16, liebchen17, hoell17, liebchen17b}, laning~\cite{menzel13, kogler15, romanczuk16, menzel16njp, reichhardt18}, motility-induced phase separation~\cite{tailleur08, palacci13, stenhammar13, buttinoni13, speck14, speck15}, swarming~\cite{yang10pre, thutupalli11, wensink12}, and active turbulence~\cite{wensink12pnas, dunkel13, heidenreich14, kaiser14, heidenreich16, lowen16, thampi16, doostmohammadi17} are observed.

In many cases, artificial self-driven particles and swimming microorganisms have to propel through complex fluids, such as polymer gels and viscoelastic microemulsions~\cite{fu07, lauga07, lauga09epl, liu11, shen11, schwarz12, zhu12, qiu14, riley14, gomez16, singh17acsnano}.
Notable examples include sperm navigation through the mammalian female reproductive tract~\cite{suarez06, kantsler14}, bacteria locomotion in biofilm matrices composed of extracellular polymeric substances~\cite{smalyukh08, flemming10}, nematode movement in soil~\cite{wallace68, ronn15}, and the motion of synthetic microswimmers and microrobots in blood vessels for targeted drug delivery applications~\cite{wang09acsnano, gao13, gao14, rao15, vikram16, hosseinidoust16}.
A wealth of fascinating behaviors emerges from the coupling between the activity of self-driven particles and the complexity of the host fluid.

Among complex fluids, liquid crystals (LCs) show states of matter,  the physical properties of which are intermediate between conventional liquids and solid crystalline states~\cite{martin72, degennes95}.
They consist, for instance, of elongated rod- or flat disc-like organic molecules that, for example, display a collective orientational order along one axis described by the so-called director~\cite{lavrentovich16}.
This simplest form of LCs are the uniaxial nematics, the rheological properties of which were characterized by many works~\cite{kneppe81, beens83, ehrentraut95,kroger97, kroger97b, negita04}.

Mathematically, the hydrodynamic coupling between the nematic director and the velocity field is, for example, described using a continuum mechanics approach based on the Leslie-Ericksen theory~\cite{leslie68, ericksen69, helfrich69, stephen74}.
An alternative formulation of the hydrodynamic behavior of LCs, based on standard conservation equations of mass, energy, momentum, and on appropriate equations for variables describing the underlying spontaneously broken symmetries,  has likewise been proposed~\cite{martin70, forster71, forster75book, pleiner96, pleiner02}.

{To measure the rheological properties of complex fluids, microrheological techniques are promising means.
These are based on the observation of either the non-driven, passive motion of probe particles inserted into the surrounding medium, or of their driven motion when subject to imposed forces.}
The  motion of passive particles immersed in a LC is quite well understood and has thoroughly been studied since many years ago~\cite{stark99, stark01review, turiv13}.
Given types of anchoring and alignment of the LC molecules on the particle surfaces elastically distort an otherwise spatially uniform alignment of the director field~\cite{cordoba16}.
The consequential elastic energy in anisotropic LCs can result in a novel class of colloidal interactions between particles~\cite{poulin97, poulin98}.
Experimentally, it has been shown that self-diffusion in a nematic LC obeys a generalized Stokes-Einstein relation with the effective diffusion coefficient along the oriented far-field director usually larger than perpendicular to it in the investigated cases~\cite{pasechnik04, hasnain06, gleeson06, mondiot12}.

The axisymmetric flow field around a sphere dragged along the director of an undisturbed, aligned nematic LC has theoretically been studied in the particular case of neglecting one viscosity coefficient~\cite{kneppe91}.
The method yields analytical expressions for the frictional drag acting on a sphere in an otherwise quiescent LC. 
A general solution for arbitrary orientation and viscosity coefficients has later been obtained numerically using a finite-difference approach~\cite{heuer92}.
More recently, closed-form analytical formulas derived from conservation laws for nematic LCs have been derived using a Fourier transform technique~\cite{gomez13, gomez-gonzalez16, gomez-phd-thesis}.
{The Stokes drag of a spherical particle in various nematic environments has further been studied by means of computer simulations~\cite{stark01, stark02, loudet04, stark02jpcm}.}

Examples for investigations on active particles in nematic LCs are given, for instance, by the swimming of motile bacteria in nematic LCs.
This type of active suspensions were themed \enquote{living liquid crystals}~\cite{zhou14}.
Experiments carried out in lyotropic chromonic nematic LCs revealed that swimming bacteria, such as \textit{E. coli}~\cite{kumar13}, \textit{B. subtilis}~\cite{zhou14, sokolov15, genkin17, genkin18}, or \textit{P. mirabilis}~\cite{mushenheim14b, mushenheim15} tend to align along the local director.
This observed behavior suggests that swimming in an anisotropic medium can conveniently be utilized as a guiding strategy to direct the motion of self-propelling active agents.
Moreover, it has been shown that fluid anisotropy can significantly alter pairwise interactions between swimming bacteria and allow transport of cargo particles along predetermined trajectories defined by the nematic director~\cite{trivedi15, sokolov15}. 
In addition, the dynamical properties of self assembly have been studied for motile bacteria~\cite{mushenheim14} and phoretically driven active particles~\cite{hernandez15}.
The effects of thermal fluctuations of the nematic director on the dynamics of a model active particle moving in a LC have further been considered theoretically and by means of computer simulations~\cite{toner16, ferreiro18}.
Meanwhile, the effect of liquid-crystalline anisotropy on the behavior of a classical Taylor swimming sheet~\cite{taylor51} undulating with small-amplitude traveling waves has been examined~\cite{krieger14, krieger15, krieger15epje,shi17}.

Recently, the dynamics of a self-propelled spherical model squirmer with a prescribed tangential slip velocity on its surface~\cite{lighthill52} has been investigated in a nematic LC using a combination of lattice Boltzmann simulations and analytical calculations~\cite{lintuvuori17}.
In the steady state of motion, it has been demonstrated that a pusher-type (extensile) swimmer will swim along the nematic director while a puller-type (contractile) swimmer will align along an axis perpendicular to the director. 
The emerging reorientation of the swimmer has been attributed to the hydrodynamic coupling between the flow field induced by the squirmer and the anisotropy described by the liquid-crystalline viscosities.
In the weakly anisotropic limit, the behavior of a general axisymmetric microswimmer described as a linear combination of higher-order singularity solutions of the Stokes flow has further been considered~\cite{kos18}.

In the present manuscript, we study theoretically the swimming behavior of a minimal model microswimmer freely moving in a viscous uniaxially anisotropic fluid.  
The axisymmetric swimmer is modeled as a sphere asymmetrically placed between two active force centers that set the surrounding fluid into motion~\cite{menzel16, hoell17}.
Both pusher- and puller-type microswimmers can be realized in this way.
We include for the fluid flow in the surrounding medium the viscosity tensor of the same anisotropic uniaxial structure as in a nematic LC~\cite{martin70, forster71, forster75book, pleiner96, pleiner02}.

Depending on the propulsion mechanism (pusher or puller), the initial orientation, and the ratio between the anisotropic viscosities, we find that the swimmer aligns parallel or perpendicular to the director, or, under some circumstances, assumes a steady intermediate orientation.
Our results indicate that the orientational behavior observed in recent experiments for different types of microswimmers in nematic LCs can be understood qualitatively already from the resulting anisotropy in the overall induced flow field surrounding the microswimmer.
Consequently, by adjusted director configurations or induced switching of the nematic  director orientation, individual microswimmers can be guided along a requested path.

The remainder of the paper is organized as follows.
In Sec.~\ref{sec:mathematischeFormulierung}, we overview the low-Reynolds-number continuum description used to characterize the dynamics of the surrounding anisotropic fluid in relation to conventional nematic LCs.
We then derive in Sec.~\ref{sec:GreenschenFunktionen} explicit expressions for the Green's function, which is the solution of the governing equations for a point-force singularity acting in the fluid domain.
In Sec.~\ref{sec:SchwimmenInEinemNematischenFluessigkristall}, we present our minimal model microswimmer and investigate its swimming behavior in the anisotropic fluid.
Concluding remarks are offered in Sec.~\ref{sec:Schluss}, and technical details are relegated to the Appendices.
We outline in Appendix~\ref{appendix:inverseFourierTransformation} the algebra leading to the derivation of the Green's function using the Fourier transformation technique.
In Appendix~\ref{appendix:deltaXZ}, we show some analytical calculations for the distances traveled by a microswimmer during reorientation.
We then quantify in Appendix~\ref{appendix:HydrodynamischeMobilitaeten} the fluid-mediated hydrodynamic interactions between colloidal particles which could serve as a basis for future investigations of the behavior of special particle-based microswimmer models~\cite{najafi04, najafi05, golestanian08, golestanian08epje, golestanian09jpcm, ledesma12, pande17, liebchen18viscotaxis, daddi18, daddi18jpcm, lowen18}.

% % % % % % % % % % % % % % % % % % % % % % % % % %

\section{Low-Reynolds-number flows in the anisotropic medium}\label{sec:mathematischeFormulierung}

Typically, due to their microscopic size, the flows of microswimmers induced in the surrounding medium are characterized by low Reynolds numbers. Thus, it is the Stokes equation that dominates the dynamics of the surrounding fluid, which we may generally write as 
\begin{equation}
\label{Stokes}
{}\nabla_j\sigma_{ij} (\R) = f_i(\vect{r})  \, . 
\end{equation}
In this expression, $\sigma_{ij}(\R) = p(\R) \delta_{ij}+\tilde{\sigma}_{ij} (\R)$ denote the components of the stress tensor, with $p(\R)$ the thermodynamic pressure field, $\delta_{ij}$ the Kronecker delta, $\vect{f}(\vect{r})$ a force density acting on the fluid, and summation over repeated indices is implied. In a simple isotropic fluid, $\bm{\tilde{\sigma}} (\R)$ introduces the effect of viscous dissipation into the dynamic equations. It contains two viscosity parameters, one associated with the shear viscosity and one with dissipation under volume changes. 
{We note that the sign convention for $\bm{\tilde{\sigma}}(\R)$ varies in the literature, and the corresponding tensor may also be found to be defined with a minus sign.}

In a uniaxially anisotropic fluid, the situation becomes more complex. We start from the hydrodynamic symmetry-based equations for a conventional nematic LC deep in the nematic phase as can be found, e.g., in Refs.~\onlinecite{martin72,forster75book,pleiner96}. The local orientation of the axis of uniaxial order is then characterized by the director field $\vect{\hat{n}}(\vect{r})$ of unit magnitude $|\vect{\hat{n}}(\vect{r})|=1$. 
In general, $\vect{\hat{n}}(\vect{r})$ may be a dynamic field, possibly changing over time as a consequence of fluid flows. Simultaneously, spatial variations of $\vect{\hat{n}}(\vect{r})$ cost energy and contribute to the stress tensor $\bm{\tilde{\sigma}} (\R)$, as do, for instance, deviations from aligning magnetic or electric fields \cite{pleiner02}, thus inducing fluid flows. 

To be able to make analytical progress below, we consider the director permanently and spatially homogeneously aligned along one global axis \cite{kneppe81,heuer92,kos18}. For instance, this could be achieved by a strong aligning homogeneous external electric (under insulating conditions) or magnetic field \cite{degennes95,pleiner96,pleiner02}. For perfect alignment of $\vect{\hat{n}} (\R)$ along the field, also the corresponding contributions to the stress tensor drop out.  
{The assumption of an undistorted, spatially homogeneous nematic director field can be substantiated for cases in which the magnitude of the so-called Ericksen number $\mathrm{Er}$ is small \cite{stark2001stokes}. Here, the Ericksen number is denoted as \cite{larson1993ericksen} $\mathrm{Er}=\gamma_1UL/K$. In this expression, $\gamma_1$ is the so-called rotational viscosity associated with pure director rotations \cite{degennes95}, $U$ is a typical speed, $L$ is a typical length scale, and $K$ is the order of magnitude of the Frank elastic coefficients associated with elastic distortions of the homogeneous director field \cite{degennes95}. Inserting as a typical size of a microswimmer $L\sim10^{-6}$~$\mathrm{m}$, as a typical speed $U\sim10^{-6}$~${\mathrm{m}}/{\mathrm{s}}$, as well as characteristic orders of magnitude of the material parameters for the commonly used liquid crystal 5CB, namely $\gamma_1\sim0.1$~$\mathrm{Pa}\,\mathrm{s}$ for the rotational viscosity \cite{kneppe1982rotational} and $K\sim10^{-11}~\mathrm{N}$ for the Frank elastic coefficients \cite{skarp1980measurements}, we obtain $\mathrm{Er}\sim10^{-2}\ll1$.} 

As a consequence, we reduce the stress tensor $\bm{\tilde{\sigma}} (\R)$ to the dissipative stress tensor associated with gradients in the fluid flow~\cite{pleiner96,pleiner02},
\begin{equation}
\tilde{\sigma}_{ij} = \sigma_{ij}^{\text{D}} = {}-\nu_{ijkl}\nabla_l v_k \, . \label{momentumEquation}
\end{equation}
Here, $\vect{v} (\R)$ is the velocity field, not writing the dependence on $\R$ explicitly any longer, and $\nu_{ijkl}$ is the material viscosity tensor of uniaxial symmetry, given by \cite{pleiner96}
\begin{eqnarray}
\nu_{ijkl} &=& \nu_2(\delta_{ik}\delta_{jl} + \delta_{il}\delta_{jk}) %\nonumber\\
%&& { }
+ 2(\nu_1+\nu_2-2\nu_3)n_in_jn_kn_l %\nonumber\\
%&& {} 
+ (\nu_3-\nu_2)(n_in_k\delta_{jl}+n_in_l\delta_{jk}+n_jn_k\delta_{il}+n_jn_l\delta_{ik}) \nonumber\\
&& {} 
+ (\nu_5-\nu_4+\nu_2)(\delta_{ij}n_kn_l+\delta_{kl}n_in_j) %\nonumber \\
%&& {} 
+(\nu_4-\nu_2)\delta_{ij}\delta_{kl} \, . \label{nu_ijkl}
\end{eqnarray}
The remaining role of the director $\vect{\hat{n}}$ is thus to render the viscosity tensor uniaxially anisotropic. From now on, we choose $\vect{\hat{n}}\parallel\vect{\hat{z}}$, yielding 
\begin{eqnarray}
-\sigma_{ij}^{\text{D}} &=& \nu_2(v_{i,j} + v_{j,i})
+ 2(\nu_1+\nu_2-2\nu_3)\delta_{iz}\delta_{jz} v_{z,z} %\nonumber\\
%&&{} 
+ (\nu_3-\nu_2)\big( \delta_{iz} \left( v_{z,j} + v_{j,z} \right) + \delta_{jz} \left( v_{z,i} + v_{i,z} \right) \big) \nonumber\\
&&{}
+ (\nu_5-\nu_4+\nu_2)\left(\delta_{ij} v_{z,z}+\delta_{iz}\delta_{jz} v_{k,k}\right) %\notag \\
%&&{}
+(\nu_4-\nu_2)\delta_{ij}v_{k,k}  \, ,
\label{eq_sigmaD}
\end{eqnarray}
where commas denote partial derivatives. 

As a next step, we include the typical assumption in related considerations of incompressible fluid flows, i.e., local volume conservation and constant density. Consequently, the continuity equation reduces to 
\begin{equation}
\nabla\cdot\vect{v}=0 \, . \label{incompressibilityEqn}
\end{equation}
Physically, it is the pressure field $p$ in Eq.~(\ref{Stokes}) that needs to guarantee this relation. Under the initial prerequisite of $\nabla\cdot\vect{v}=0$, it thus needs to ensure $\partial_t\nabla\cdot\vect{v}=0$ at all later times \cite{pleiner96,pleiner02}. 
In contrast to an isotropic fluid, where this implies the condition $\Delta p=0$ for the pressure (at least away from any singularities), under our assumptions, one obtains the condition \cite{pleiner96,pleiner02}
\begin{equation}
\label{eq_p}
\Delta p = 2(\nu_1+\nu_2-2\nu_3) v_{z,zzz} + (2\nu_3-\nu_2+\nu_5-\nu_4)\Delta v_{z,z} \, 
\end{equation} 
that the thermodynamic pressure needs to satisfy. 

Next, the pressure is redefined as \cite{pleiner02}
\begin{equation}
\label{p_shift}
p^\prime = p - (\nu_5-\nu_4+\nu_2) v_{z,z} \, . 
\end{equation}
That is, instead of including the isotropic contribution $(\nu_5-\nu_4+\nu_2)\delta_{ij}(\partial_zv_z)$ into the viscous stress tensor $\sigma^{\text{D}}_{ij}$ [in the second-to-last term in Eq.~(\ref{eq_sigmaD})], it is drawn forward into the pressure in Eq.~\eqref{Stokes}. 
Then, combining Eqs.~(\ref{Stokes}) and \eqref{eq_sigmaD}--%, (\ref{incompressibilityEqn}), and 
\eqref{p_shift}, we obtain
\begin{equation}
\label{Stokes_prime}
\nabla_i p^\prime + \nabla_j\sigma_{ij}^{\prime\text{D}} = f_i  \, ,
\end{equation}
where
\begin{eqnarray}
-\sigma_{ij}^{\prime\text{D}} &=& \nu_2( v_{i,j} + v_{j,i}) %\nonumber\\
%&&{}
+ 2(\nu_1+\nu_2-2\nu_3)\delta_{iz}\delta_{jz} v_{z,z} %\nonumber\\
%&&{}
+ (\nu_3-\nu_2)\big(\delta_{iz}\left( v_{z,j} + v_{j,z} \right) + \delta_{jz} \left( v_{z,i} + v_{i,z} \right) \big) \,  \qquad 
\label{eq_sigmaD_prime}
\end{eqnarray}
and $p'$ needs to satisfy %the modified version of Eq.~(\ref{eq_p}) to yield, 
\begin{equation}
\label{eq_p_prime}
\Delta p^\prime = 2(\nu_1+\nu_2-2\nu_3) v_{z,zzz} + 2(\nu_3-\nu_2)\Delta v_{z,z} \, .
\end{equation}
The benefit of this transformation is that in the incompressible case the number of involved viscosities can thus be formally reduced to three, namely $\nu_1$, $\nu_2$, and $\nu_3$. Obviously, it is the viscosity $\nu_2$ that leads the term that is also present in the description of an isotropic liquid. Switching back to the physical pressure field involves the combination of viscosities $\nu_5-\nu_4$. 
It was demonstrated in Refs.~\onlinecite{pleiner96,pleiner02} that working with the redefined pressure, the present formalism is compatible with the original one by Leslie-Ericksen~\cite{leslie68, ericksen69, degennes95}.
For convenience, we repeat here the relations between the
Leslie-Ericksen parameters $\alpha_1$, $\alpha_2$, $\alpha_3$,
$\alpha_4$, $\alpha_5$, $\alpha_6$, $\gamma_1$, as well as $\gamma_2$
and the parameters used here, as listed in Ref.~\onlinecite{pleiner96}:
$\alpha_1=2 \left(\nu_1+\nu_2-2\nu_3\right)-\gamma_1 \lambda^2, \alpha_2=-\gamma_1(1+\lambda)/2, \alpha_3=\gamma_1(1-\lambda)/2, \alpha_4=2\nu_2, \alpha_5=2(\nu_3-\nu_2) + \gamma_1\lambda (\lambda+1)/2, \alpha_6=2(\nu_3-\nu_2)+\gamma_1\lambda (\lambda-1)/2$,
and $\gamma_2=-\gamma_1 \lambda$, where $\lambda$ is the parameter of flow alignment
\cite{pleiner96}.
If one is interested in the thermodynamic pressure, one would at the
end need to switch back to the pressure field~$p$~\cite{pleiner02}.

{For convenience, we here abbreviate
\begin{equation}
	\bar{\nu} %&
	= 2(\nu_1+\nu_2-2\nu_3) \, . \label{relation_with_Leslie-Ericksen_1} %\\
\end{equation}
Later in this work, an expansion around $\bar{\nu}=0$ will be performed. As can be inferred, e.g., from Eq.~(\ref{eq_sigmaD}), setting $\bar{\nu}=0$ neglects viscous forces parallel to the director on surfaces with their normal along the director, caused by longitudinal variations of the velocity along the director.}

% % % % % % % % % % % % % % % % % % % % % %

\section{Green's function}\label{sec:GreenschenFunktionen}

\subsection{Solution in Fourier space}

Having outlined the hydrodynamic equations governing the dynamics of our uniaxial anisotropic fluid at low Reynolds numbers, we now derive explicit expressions for the Green's function representing the solution for the velocity field~$\vect{v} (\R) $ at position~$\R$, due to a point force density $\vect{f}(\R) = \vect{F} \delta (\R-\R_0)$ acting on the fluid domain at position~$\R_0$.
Then,
\begin{equation}
	\vect{v} (\R) = \boldsymbol{\G} (\R - \R_0) \cdot \vect{F}  \, .
	\label{GreenDefinition}
\end{equation}
For an isotropic fluid of dynamic viscosity~$\eta$, the Green's function is given by the Oseen tensor~\cite{happel12, leal80}, namely, $\G_{ij}=(8\pi\eta R)^{-1}(\delta_{ij} + R_i R_j/R^2)$, where $\boldsymbol{R}=\R-\R_0$ and $R=|\vect{R}|$.
The corresponding solution for the pressure is $p = \boldsymbol{\mathcal{P}} \cdot \vect{F}$ where $\mathcal{P}_j = R_j/4\pi R^3$.
Thanks to the linearity of the hydrodynamic equations, the solution for an arbitrary force distribution can readily be determined by the superposition principle~\cite{barton89}.

The momentum equation for an anisotropic fluid given in a vectorial form by~Eq.~\eqref{Stokes_prime} can be projected onto the Cartesian coordinate basis to obtain
\begin{subequations}\label{momentumEqs}
	\begin{align}
	   -P_{,x} + \nu_3 v_{x,zz} + \nu_2 \left( v_{x,xx} + v_{x,yy} \right) + F_x \, \delta (\vect{R} ) = 0 \, , \label{momentum_X} \\
	   -P_{,y} + \nu_3 v_{y,zz} + \nu_2 \left( v_{y,xx} + v_{y,yy} \right) + F_y \, \delta (\vect{R} ) = 0 \, , \label{momentum_Y} \\
	   -P_{,z} + K v_{z,zz} + \nu_3 \left( v_{z,xx} + v_{z,yy} \right) + F_z \, \delta (\vect{R} ) = 0 \, , \label{momentum_Z}
	\end{align}
\end{subequations}
where, for convenience, we have once more redefined the pressure variable as
\begin{equation}
    P = p^\prime- (\nu_3-\nu_2) v_{z,z} \, , 
    %= p - (\nu_5-\nu_4+\nu_3) v_{z,z} 
\end{equation}
in addition to the viscosity coefficient 
\begin{equation}
 K = 2\nu_1 + \nu_2 - 2\nu_3 \, .
\end{equation}

Solving Eqs.~\eqref{momentumEqs} for the velocity and pressure fields can conveniently be performed using the Fourier transform technique~\cite{bracewell99}.
At distances far away from~$\R_0$, we assume that the flow fields decay to zero, so that the Fourier transforms are well defined. 
We calculate the 3D (forward) Fourier transform of a function $g(\R)$ given in real space as
\begin{equation}
\mathscr{F} \left\{ g(\R) \right\} = \tilde{g} (\K) = \int_{\mathbb{R}^3} g(\R) \, e^{-i \K \cdot \R} \, \Intd \R \, , \label{FourierForward}
\end{equation}
together with the inverse Fourier transform
\begin{equation}
\mathscr{F}^{-1} \left\{ \tilde{g}(\K) \right\} = {g} (\R) = \frac{1}{(2\pi)^3} \int_{\mathbb{R}^3} \tilde{g} (\K) \, e^{i \K \cdot \R} \, \Intd \K \, , \label{FourierInverse}
\end{equation}
where $\K = (k_x, k_y, k_z)$ is the wavevector that sets the coordinates in Fourier space.

% Refs by bickel checked and are ok

It turns out to be adequate to employ the orthogonal coordinate system previously introduced by Bickel \cite{bickel07, bickel06} in which the Fourier-transformed vector components in the plane perpendicular to the director are decomposed into longitudinal and transverse components~\cite{Bickel10, daddi16, daddi18epje}. 
For a given vector quantity~$\tilde{\vect{Q}}$, the components of which in the Cartesian coordinate basis are $(\tilde{Q}_x, \tilde{Q}_y, \tilde{Q}_z)$, the corresponding components in the new orthogonal basis formed by the unit vectors $\{\vect{e}_l, \vect{e}_t,\vect{e}_z \}$ are given implicitly by the orthogonal transformation
\begin{equation}  \label{transformation}
\left( 
      \begin{array}{c}
      \tilde{Q}_x \\
      \tilde{Q}_y \\
      \tilde{Q}_z
      \end{array}
\right)
=
\frac{1}{k_{\perp}}
\left( 
      \begin{array}{ccc}
      k_x & k_y & 0 \\
      k_y & -k_x & 0 \\
      0 & 0 & k_\perp
      \end{array}
\right)
\left( 
      \begin{array}{c}
      \tilde{Q}_l \\
      \tilde{Q}_t \\
      \tilde{Q}_z
      \end{array}
\right) \, ,
\end{equation}
where $\tilde{Q}_l$ and $\tilde{Q}_t$ refer to the longitudinal and transverse vector components, respectively.
Moreover, $k_{\perp} = \sqrt{k_x^2 + k_y^2}$. 
We note that the components along the $\vect{\hat{z}}$ direction parallel to the director are not affected by this transformation.
Transforming the momentum equations stated by Eqs.~\eqref{momentumEqs} to Fourier space and projecting the resulting equations onto the new orthogonal basis, we obtain
\begin{subequations} \label{momentum_LTZ}
	\begin{align}
	 i k_\perp \tP &= - \left( \nu_3 k_\parallel^2 + \nu_2 k_\perp^2 \right) \tv_l + {F}_l \, , \label{momentum_L} \\
	  0 &= - \left( \nu_3 k_\parallel^2 + \nu_2 k_\perp^2 \right) \tv_t + {F}_t \, , \label{momentum_T} \\
	 i k_\parallel \tP &= - \left( K k_\parallel^2+\nu_3 k_\perp^2 \right) \tv_z + {F}_z \, ,  \label{momentum_Z_trans}
	\end{align}
\end{subequations}
where we have used the notation $k_\parallel := k_z$.
In addition, by transforming Eq.~\eqref{incompressibilityEqn} to Fourier space, a direct relation between the components $\tv_l$ and $\tv_z$ can be established.
Specifically, 
\begin{equation}
  k_\perp \tv_l + k_\parallel \tv_z = 0 \, . \label{incompressibilityEqn_New}
\end{equation}

It can be noted from Eq.~\eqref{momentum_T} that the transverse component $\tv_t$ is independent of the pressure variable and can be separated from the longitudinal and normal velocity components.
Solving this equation for $\tv_t$ yields
\begin{equation}\label{vt_sol}
  \tv_t = \tG_{tt} F_t = \frac{{F}_t}{\nu_3 k_\parallel^2 + \nu_2 k_\perp^2 } \, .
\end{equation}

Furthermore, by combining Eqs.~\eqref{momentum_L} and \eqref{momentum_Z_trans}, the pressure variable~$\tilde{P}$ can be eliminated.
Upon making use of Eq.~\eqref{incompressibilityEqn_New}, the solutions for the normal and longitudinal velocities read
\begin{subequations}\label{vz_vl_sol}
	\begin{align}
	  \tv_z &= \tG_{zz}F_z + \tG_{zl}F_l = \frac{k_\perp^2 {F}_z - k_\perp k_\parallel {F}_l}{\nu_3 k^4 + \bar{\nu} k_\parallel^2 k_\perp^2} \, , \label{normalVelocity} \\
	  \tv_l &= \tG_{ll}F_l + \tG_{lz}F_z = \frac{k_\parallel^2 {F}_l - k_\perp k_\parallel {F}_z }{\nu_3 k^4 + \bar{\nu} k_\parallel^2 k_\perp^2}  \, , \label{longitudinalVelocity}
	\end{align}
\end{subequations}
where we have {employed the abbreviation $\bar{\nu}$ 
defined in} Eq.~\eqref{relation_with_Leslie-Ericksen_1}.

Based on Eqs.~\eqref{vt_sol} and \eqref{vz_vl_sol}, the Green's tensor in the new vector basis $\{\vect{e}_l, \vect{e}_t,\vect{e}_z \}$ is a function of $k_\parallel$ and $k_\perp$ only.
It can be expressed in a matrix form as
\begin{equation}
 \boldsymbol{\tG} = 
		\begin{pmatrix} 
		     \cfrac{k_\parallel^2 }{\nu_3 k^4 + \bar{\nu} k_\parallel^2 k_\perp^2} & 0 & \cfrac{ - k_\perp k_\parallel }{\nu_3 k^4 + \bar{\nu} k_\parallel^2 k_\perp^2} \\
		     0 &  \cfrac{1}{\nu_3 k_\parallel^2 + \nu_2 k_\perp^2 } & 0 \\
		     \cfrac{ - k_\perp k_\parallel }{\nu_3 k^4 + \bar{\nu} k_\parallel^2 k_\perp^2} & 0 & \cfrac{k_\perp^2}{\nu_3 k^4 + \bar{\nu} k_\parallel^2 k_\perp^2}
		\end{pmatrix}  .
		\label{GreenTensor_Fourier_ltz}
\end{equation}

% use pmatrix to get the equation numbering at the same line! don't use array

The components of this Green's function in the original Cartesian basis in Fourier space are obtained by means of a standard change of basis~\cite{trefethen97}.
Following Eq.~\eqref{transformation}, we obtain
\begin{subequations}\label{GreenFunctions_Fourier_xyz}
	\begin{align}
	\tG_{xx} &= \tG_{tt}  \sin^2\varphi_k + \tG_{ll}  \cos^2 \varphi_k \, , \label{G_xx} \\
	\tG_{yy} &= \tG_{tt}  \cos^2\varphi_k + \tG_{ll}  \sin^2 \varphi_k \, , \label{G_yy} \\
	\tG_{xy} &= \left( \tG_{ll} - \tG_{tt} \right) \cos\varphi_k \sin\varphi_k \, , \label{G_xy} \\
	\tG_{xz} &= \tG_{lz} \cos\varphi_k \, , \label{G_xz} \\
	\tG_{yz} &= \tG_{lz} \sin\varphi_k \, , \label{G_yz}
	\end{align}
\end{subequations}  
% where $\varphi_k = \arctan (k_y/k_x) \in [0,2\pi]$ denotes the azimuthal angle in Fourier space.
where the angle $\varphi_k$ follows from the representation of the wavevector~$\vect{k}$ in Fourier space in spherical coordinates as
\begin{equation}
	\vect{k} = k \left( 
		 	\begin{array}{c}
		 	\sin\vartheta_k \cos\varphi_k \\	
		 	 \sin\vartheta_k \sin\varphi_k \\
		 	\cos\vartheta_k
		 	\end{array}
		 	 \right) \, .
\end{equation}
Moreover, $\tG_{yx} = \tG_{xy}$, $\tG_{zx} = \tG_{xz}$, and $\tG_{zy} = \tG_{yz}$ as required by the symmetry of the Green's tensor in an unbounded domain.

Finally, the redefined pressure variable associated with this flow field can be calculated from Eqs.~\eqref{momentum_LTZ} as
\begin{equation}
	\tilde{P} = -i \, \frac{k_\parallel \left( \nu_2 k_\perp^2+\nu_3 k_\parallel^2 \right) F_z 
						+k_\perp \left( K k_\parallel^2+\nu_3 k_\perp^2 \right) F_l}{\nu_3 k^4+\bar{\nu} k_\parallel^2 k_\perp^2} \, .
\end{equation}

Expressions of the Green's function for the velocity and pressure fields in real space are obtained via inverse Fourier transform according to Eq.~\eqref{FourierInverse}.
Since the goal of the present work is to study the behavior of a model microswimmer in an anisotropic medium, we confine ourselves to analytical expressions for the velocity field in real space.

\subsection{Solution in real space}

The inverse Fourier transform of Eqs.~\eqref{GreenFunctions_Fourier_xyz} according to Eq.~\eqref{FourierInverse} is straightforward, though laborious, and thus is shifted to Appendix~\ref{appendix:inverseFourierTransformation}.
As shown there, the final expressions of the Green's function for the velocity field in real space, see Eq.~\eqref{GreenDefinition}, can conveniently be expressed in terms of the following set of convergent definite integrals,
%\begin{widetext}
\begin{subequations}\label{greenFctsRealSpace_Main}
	\begin{eqnarray}
	 \G_{zz} &=& \frac{1}{2\pi^2 R} 
	 \int_{0}^{1} \frac{(1-\sz^2 Q^2) \, \Intd Q}{\big( \nu_3+\bar{\nu} \sz^2Q^2(1-\sz^2Q^2) \big) \sqrt{1-Q^2}} \, , \label{mu_zz_P_Main} \\
	 \G_{xx} &=& \frac{1}{4\pi^2 R} \int_0^{1} 
	      \bigg(\frac{\sz^2 Q^2 \Gamma_{-} }{ \nu_3+\bar{\nu} \sz^2 Q^2(1-\sz^2 Q^2)} %\notag   \\
	         %&&{}
	         + \frac{\Gamma_{+} }{ (\nu_3-\nu_2)\sz^2 Q^2+\nu_2 } \bigg)  \, \Intd Q \, , \label{mu_xx_P_Main} \\
	 \G_{xz} &=& \frac{\cos\varphi}{2\pi^2 R} 
	  \int_0^{1} \frac{\cz \sz Q^2 \, \Intd Q}{\big( \nu_3+\bar{\nu} \sz^2 Q^2(1-\sz^2 Q^2) \big) \sqrt{1-Q^2}} \, , \label{mu_xz_P_Main} \\
	 \G_{xy} &=& \frac{\sin (2\varphi) }{4\pi^2 R} 
	      \int_0^{1} \bigg( \frac{1}{(\nu_3-\nu_2)\sz^2 Q^2+\nu_2} %\notag \\
	      %&&{}
	      -\frac{\sz^2 Q^2}{\nu_3+\bar{\nu} \sz^2 Q^2(1-\sz^2 Q^2)}  \bigg) 
	                \frac{1-(2-\sz^2)Q^2}{(1-\sz^2 Q^2)\sqrt{1-Q^2} } \, \Intd Q \, ,
	                \label{mu_xy_Pair_Main}
	\end{eqnarray}
\end{subequations}
%\end{widetext}
where, again, $R := |\vect{R}|$ denotes the radial distance from the singularity.
In addition, $\varphi$ and $\vartheta$ denote the azimuthal and polar angles, respectively, such that
\begin{equation}
	\vect{R} = \R-\R_0 
	= \left( 
	\begin{array}{c}
	x-x_0 \\
	y-y_0 \\
	z-z_0
	\end{array}
	 \right) 
	 = R \left( 
	 	\begin{array}{c}
	 	\sin\vartheta \cos\varphi \\
	 	 \sin\vartheta\sin\varphi\\
	 	\cos\vartheta
	 	\end{array}
	 	 \right)   \, .
\end{equation}
We further define the shorthand notations $\sz:= \sin\vartheta$ and $\cz:=\cos\vartheta$.
Moreover,
\begin{equation}
 \Gamma_{\pm} = \left( 1 \pm \frac{1-(2-\sz^2)Q^2}{1-\sz^2Q^2} \, \cos (2\varphi) \right) 
 \frac{1}{\sqrt{1-Q^2}} \, .
\end{equation}

As already mentioned, the remaining five components can be determined by using the symmetry property of the Green's tensor in an unbounded medium, such that $\G_{zx} = \G_{xz}$, $\G_{zy} = \G_{yz}$, and $\G_{yx} = \G_{xy}$.
The components $\G_{yy}$ and $\G_{yz}$ are determined by, respectively, substituting~$\varphi$ by $\varphi - \pi/2$ in the expressions of $\G_{xx}$ and $\G_{xz}$ given above.

For an accurate numerical evaluation of the Green's function, it is essential to remove the singularity at $Q=1$.
This can adequately be achieved by making use of the change of variable $Q=\sin\xi$, and thus $\Intd Q/\sqrt{1-Q^2}=\Intd \xi$, leading to well-behaved integrals for~$\xi$ between 0 and $\pi/2$.
An exact analytical calculation of these integrals is possible only
under some special conditions, notably when $\bar{\nu}=0$.

The Green's function derived in this section serves as a basis for the assessment of the swimming behavior in an anisotropic medium as detailed in the next section.

% % % % % % % % % % % % % % % % % % % % % % %

\section{Swimming in a nematic liquid crystal}\label{sec:SchwimmenInEinemNematischenFluessigkristall}

\begin{figure}
\begin{center}
	\includegraphics[scale=0.6]{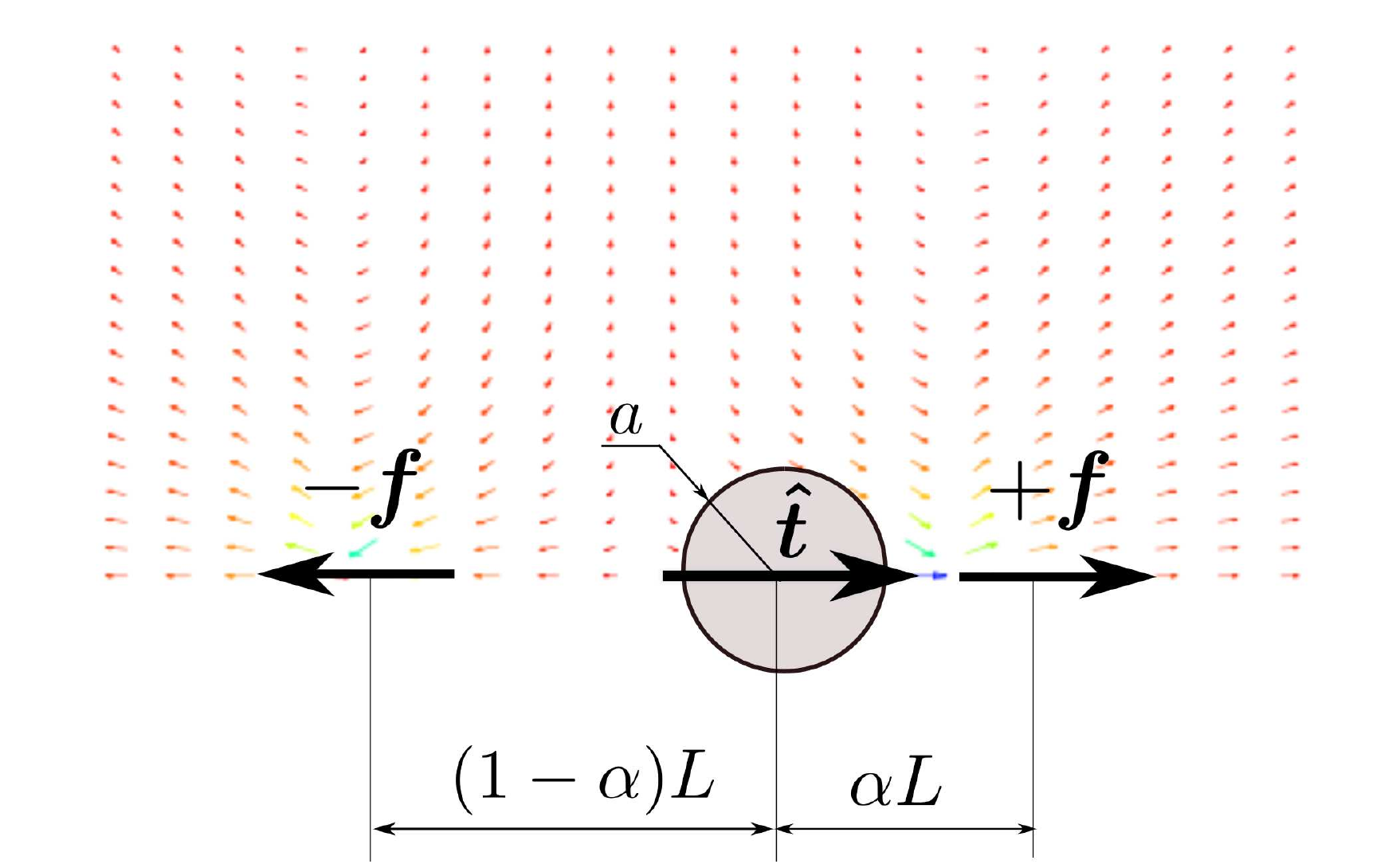}
\end{center}
%\vspace{-1cm}
\caption{(Color online) Illustration of the active model microswimmer. 
The swimmer is modeled as a sphere of hydrodynamic radius~$a$ with no-slip surface condition, subject to hydrodynamic drag.
Two point force centers exert active forces $+\vect{f}$ and $-\vect{f}$ separated a distance~$L$ from each other.
They are asymmetrically disposed with respect to the swimmer body.
Then, the resulting self-generated flow field leads to propulsion by transporting the sphere.
The positions of the force centers relative to the swimmer body are set by the parameter $\alpha$ that takes values between 0 and 1/2.
The unit vector~$\vect{\hat{t}}$ points toward the nearer force center, along the swimmer axis.
A quiver plot of the self-induced velocity field in the surrounding fluid is schematically shown for the upper half of the fluid domain in the isotropic limit for~$\nu_1=\nu_2=\nu_3$.
The microswimmer displayed here is referred to as a pusher because it pushes out the fluid along the swimming axis.
By directing the two point forces toward the spherical body, the swimmer is known as a puller because it pulls the fluid inward along the swimming path.
}
\label{modelSwimmer}
\end{figure}

We consider a minimal active model microswimmer~\cite{menzel16, hoell17} composed of a rigid sphere of radius~$a$ subject to hydrodynamic drag.
Self-propulsion is achieved through two oppositely aligned active forces $\vect{f}$ and $-\vect{f}$ oriented along the symmetry axis of the swimmer, as schematically illustrated in Fig.~\ref{modelSwimmer}.
Both puller- and pusher-type microswimmers can be modeled by directing the forces toward or away from the swimmer body, respectively.
The two force centers are separated by a distance~$L$ from each other and are asymmetrically disposed with respect to the sphere center.
This asymmetry is set through the parameter~$\alpha$ that takes values between 0 and 1/2.
The orientation of the swimmer is described by the unit vector~$\vect{\hat{t}}$ pointing into the direction of the nearer force center.
Several variants of this model can be found in the literature; see, e.g., Refs.~\onlinecite{simha02, hatwalne04, adhyapak17, degraaf16}.
Denoting by~$\R_\mathrm{S}$ the center position of the sphere, the positions of the force centers are
\begin{subequations}
	\begin{align}
		\R_+ &= \R_\mathrm{S} + \alpha L \, \vect{\hat{t}} \, , \\
		\R_- &= \R_\mathrm{S} - (1-\alpha) L \, \vect{\hat{t}} \, .
	\end{align}
\end{subequations}

Without loss of generality, we only consider in the following motion in the plane $y=0$.

\subsection{Swimming behavior for $\bar{\nu}=0$}\label{subsection:OhneAlpha1}

We begin our analysis with the particular case of~$\bar{\nu}=0$.
Exact analytical expressions for the Green's function can be obtained in this situation.
Integrating Eqs.~\eqref{greenFctsRealSpace_Main} for $y=0$ yields
\begin{subequations} \label{greenFctsLambdaZero}
	\begin{align}
		\G_{zz} &= \frac{1}{8\pi\nu_3 R} \left( 2-\sz^2 \right) \, , \\ 
		\G_{xx} &= \frac{1}{8\pi\nu_3 R} \left(  1+\sz^2
		         + \frac{2}{\sz^2} 
		         \left( \sqrt{\cz^2 + \frac{\sz^2}{E}}-1 \right) \right) \, , \label{Gxx_lambda0}\\
		\G_{xz} &= \frac{\cz \sz \cos\varphi}{8\pi\nu_3 R} \, ,
	\end{align}
\end{subequations}
and $\G_{xy}=0$, where we have defined the viscosity ratio
\begin{equation}
	E = \frac{\nu_2}{\nu_3} \, ,
\end{equation}
noting that $E \ge 0$~\cite{forster71}.
Remarkably, the anisotropy enters only through the~$xx$ component.
We further recover the Oseen tensor in the isotropic limit, for which $E=1$.

For a pusher, the flow velocity field at position~$\R$ induced by the swimmer is obtained by superimposing the flow fields due to each of the two active forces.
Specifically,
\begin{equation}
	\vect{v} (\R) = \big( \vect{\G} (\R-\R_+) - \vect{\G} (\R-\R_-) \big) \cdot \vect{f} \, .
	\label{selfInducedFlowFieldSwimmer}
\end{equation}

The flow field induced by a puller is given by the same expression with $\vect{f}$ of opposite sign. 
In the far-field limit $|\R - \R_\mathrm{S} | \gg L$, the leading-order term possesses a force-dipolar flow structure that decays as~$|\R - \R_\mathrm{S} | ^{-2}$.

Due to the complexity of the Green's function, we confine ourselves to the lowest order in the \Faxen~laws.
This implies not too large $a/(\alpha L)$ and $a/((1-\alpha)L)$ for our analysis to be valid.
The self-induced translational velocity~$\vect{V}$ and rotational velocity~$\boldsymbol{\Omega}$ of the spherical body and thus of the swimmer are
\begin{subequations}\label{defenitionVOmega}
	\begin{align}
		\vect{V} &= \vect{v}(\R_\mathrm{S}) 
		%+  \bigO \left( \big. a^2 \bNabla^2 \vect{v} \big|_{\R_0} \right) 
		\, , \\
		\vect{\Omega} &= \frac{1}{2} \, \bNabla \times \vect{v}(\R_\mathrm{S}) \, .
		%+ \bigO \left( \big. a^2 \bNabla^2 \left(\bNabla \times \vect{v} \right)\big|_{\R_0} \right) \, .
	\end{align}
\end{subequations}

We next define for convenience $\psi=\pi/2-\vartheta$ to denote the orientation angle relative to the horizontal direction, such that $\psi \in [-\pi/2,\pi/2]$.
It follows from Eqs.~\eqref{greenFctsLambdaZero} through \eqref{defenitionVOmega} that the non-vanishing components of the swimming velocities and rotation rate are given by
\begin{subequations}\label{swimmingVelocities}
	\begin{align}
		V_x &= V_0 \cos\varphi\cos\psi
		\left( \frac{w}{\cos^2\psi} -\tan^2\psi \right) \, , \label{Vx} \\
		V_z &= V_0 \sin\psi \, , \label{Vz} \\
		\Omega_y &= \Omega_0 \cos\varphi
		\tan\psi \left( 1- \frac{1}{w} \right) \, , \label{OmegaY}
	\end{align}
\end{subequations}
where
\begin{eqnarray}
	V_0 = \frac{f(1-2\alpha)}{4\pi\nu_3 L\alpha (1-\alpha)} 
\end{eqnarray}
denotes the swimming speed in an isotropic medium, for which $E=1$.
Moreover,
\begin{subequations}
	\begin{align}
		w &= \sqrt{\sin^2\psi+\frac{\cos^2\psi}{E}} \, , \label{defW} \\
		\Omega_0 &= \frac{f(1-2\alpha+2\alpha^2)}{8\pi\nu_3 L^2 \alpha^2 (1-\alpha)^2} \, . \label{Omega_0}
	\end{align}
\end{subequations}
We note that $\Omega_0/f>0$.
In addition, it can clearly be seen that no net self-induced motion occurs when the swimmer body is symmetrically located between the two force centers, i.e., for $\alpha=1/2$.
The swimmer in this configuration only pumps the fluid and is termed \enquote{shaker}.

\begin{figure}
\begin{center}
	\includegraphics[scale=1]{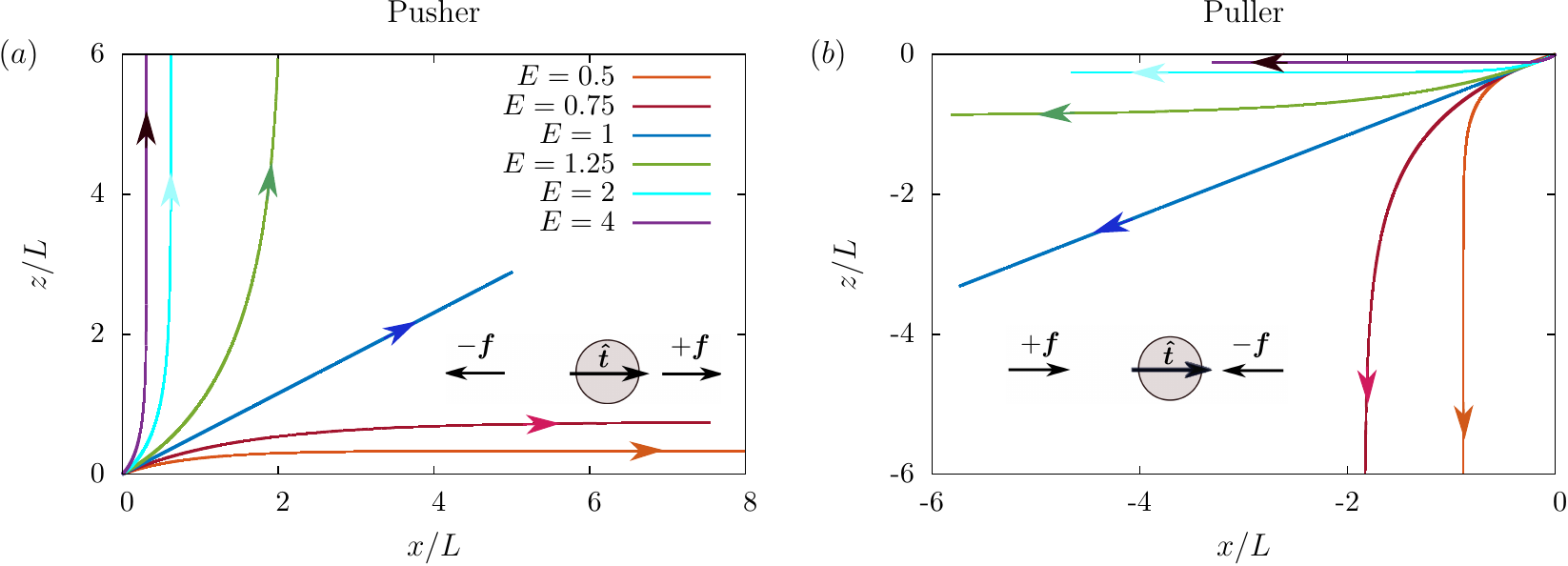}
\end{center}
\caption{(Color online) Swimming trajectories in the $(x,z)$-plane for $(a)$ pusher- and $(b)$ puller-type swimmers for a vertical director orientation along the $z$-axis and various values of the viscosity ratio~$E=\nu_2/\nu_3$. 
The swimmer is released from the origin with an initial inclination $\psi_0 = \pi/6$.
Arrows indicate the direction of time evolution.
Here, $\alpha=2/5$ and $\bar{\nu}=0$.}
\label{SwimmerTraj}
\end{figure}

In order to illustrate from the above equations the effect of the anisotropy on the swimming behavior, we consider a small deviation from the isotropic values.
By performing a Taylor expansion of the rotation rate about $E=1$, the leading-order term reads
\begin{equation}
	\Omega_y \sim  \frac{\Omega_0}{4} \left(1-E\right) \cos\varphi\sin\left(2\psi\right)  \, .
\end{equation}

Then, for a pusher-type swimmer $(\Omega_0>0)$ and for $E<1$, the orientation $\psi=0$ $(\cos\varphi = \pm 1)$ is a stable fixed point.
Accordingly, the swimmer aligns in the steady limit perpendicular to the director and moves at a speed $V_\perp =  V_0/\sqrt{E}$.
In contrast to that, the orientation~$\psi=\pm \pi/2$ is a stable fixed point for $E>1$.
In this situation, the swimmer aligns parallel to the director, and swims at a speed $V_\parallel = V_0$.
The opposite behavior is observed for a puller-type swimmer $(\Omega_0<0)$ where the alignment occurs parallel to the director for $E<1$ and perpendicular to the director for $E>1$.
In an isotropic fluid $(E=1)$, the rotation rate vanishes.
Accordingly, the swimmer maintains a constant orientation and swims along a straight trajectory.

In Fig.~\ref{SwimmerTraj}, we illustrate for various values of~$E$ exemplary swimming trajectories in the $(x,z)$-plane for $(a)$ pusher- and $(b)$ puller-type microswimmers in the particular situation of~$\bar{\nu}=0$.
The swimmer is initially released from the origin of the coordinate system with an orientation $\psi_0=\pi/6$ relative to the horizontal.
Here, we set $\alpha = 2/5$ for the swimmer asymmetry.
Results for six values of the viscosity ratio~$E$ are shown, which span a wide range of values for actual nematic liquid crystals.
In fact, we can calculate analytically from the initial conditions the overall horizontal and vertical distance covered until complete alignment parallel and perpendicular to the director is achieved, respectively.
The expressions that we found are listed in Appendix~\ref{appendix:deltaXZ}.

To gain more insight into the effect of~$E$ on the swimming behavior, we define at this point the effective viscosities associated with the motion of a particle parallel and perpendicular to the director.
In analogy to Stokes' law, we define the effective viscosities as $\etaEff_{\parallel, \perp} = 1/(6\pi a \mu_{\parallel, \perp} )$, respectively, where $\mu_{\parallel, \perp}$ stands for the hydrodynamic mobility function (c.f.\@ Appendix~\ref{appendix:HydrodynamischeMobilitaeten} for their derivation).
For $\bar{\nu}=0$, we obtain
\begin{equation}
	\frac{\etaEff_\perp}{\etaEff_\parallel} =
	\frac{4}{1 + \cfrac{3 \arctan \left(\sqrt{\frac{1}{E}-1}\right)}{\sqrt{E(1-E)}} } \, .
\end{equation}
Performing a Taylor expansion about $E=1$ to leading order yields
\begin{equation}
	\frac{\etaEff_\perp}{\etaEff_\parallel} \sim 1+\frac{1}{2} \left(E-1\right) \, .
	\label{etaEff_perturvative}
\end{equation}

For common nematic LCs, such as 5CB and MBBA, the ratio of effective viscosities is $\etaEff_\perp > \etaEff_\parallel$ (and thus $E>1$) as observed in experiments~\cite{loudet04, gleeson06, mondiot12} and in computer simulations~\cite{stark01, lintuvuori10}.
This means a higher mobility along than perpendicular to the director.
In our system, this implies that a pusher-type swimmer aligns with the director whereas a puller tends to swim in the perpendicular direction.
Such a behavior is in agreement with the theoretical predictions and lattice Boltzmann simulations of a squirmer model~\cite{lintuvuori17}, and also with the alignment dynamics of pusher-type bacteria observed in recent experiments~\cite{kumar13,zhou14, sokolov15, genkin17, genkin18, mushenheim14b, mushenheim15}.

For $E<1$ in Eq.~\eqref{etaEff_perturvative}, the ratio of effective viscosities is $\etaEff_\perp < \etaEff_\parallel$.
This situation is less common but might in practice be encountered in discotic nematics or in corresponding lyotropic micellar LCs~\cite{mondiot12}. 
In this case, our pusher tends to align along the direction perpendicular to the director, whereas the puller tends to align parallel to the director.

\begin{figure*}
	 \centering
	 \includegraphics[scale=0.95]{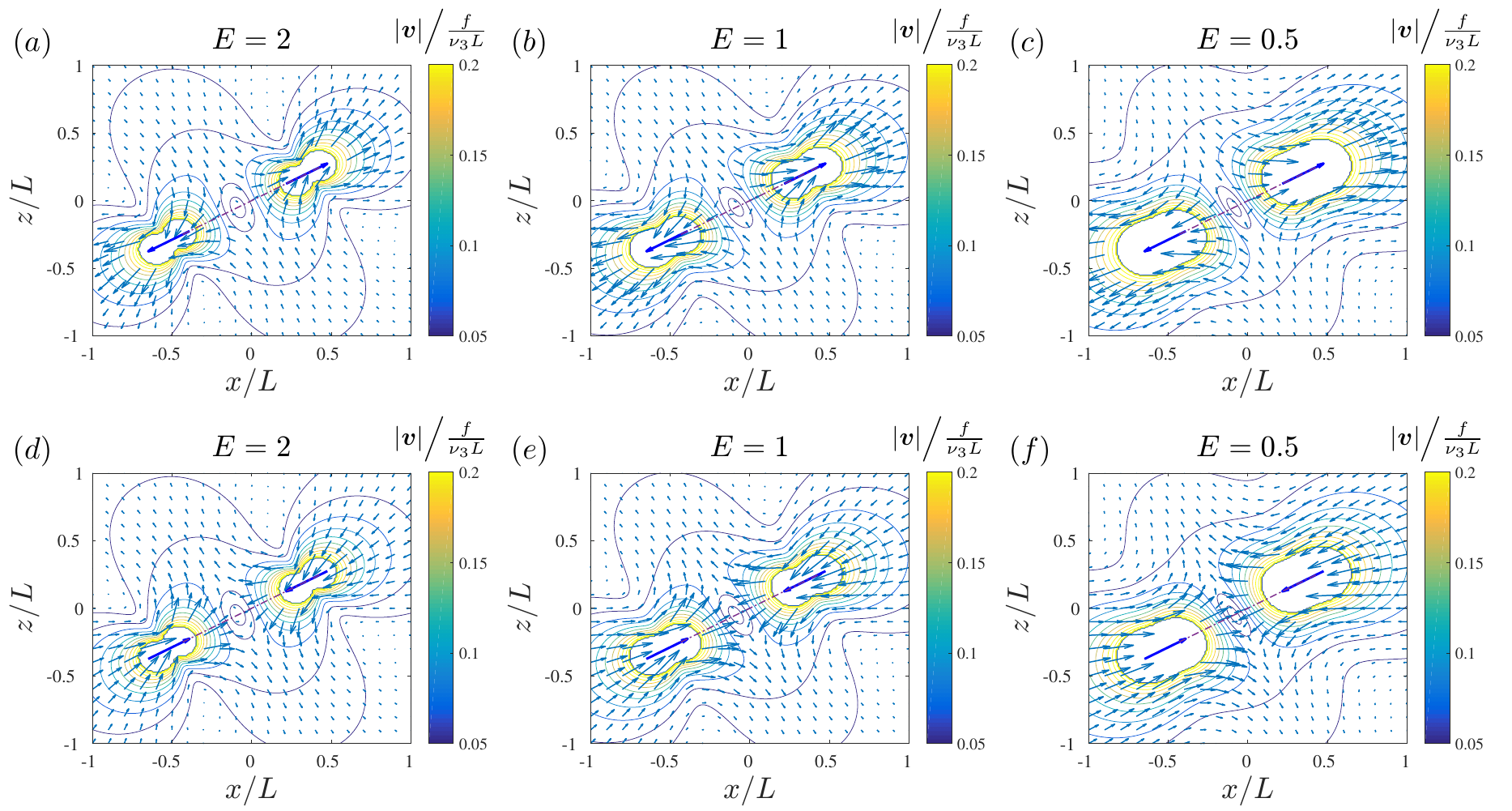}
	    \caption{(Color online) Quiver plots of the velocity field induced by a pusher- [subfigures $(a)$, $(b)$, and $(c)$] and a puller-type [subfigures $(d)$, $(e)$, and $(f)$] microswimmer in a uniaxially anisotropic fluid, e.g., an aligned nematic LC, for various values of the viscosity ratio~$E=\nu_2/\nu_3$, while $\bar{\nu}=0$. 
	    The dashed-dotted lines connecting the force centers are plotted to indicate the orientation of the swimmer.
	    Here, we set an inclination $\psi=\pi/6$ and a swimmer asymmetry $\alpha=2/5$. 
	    The color bars show the magnitude of the flow velocity scaled by $f/(\nu_3 L)$. 
	    Spherical swimmer bodies are not shown here for clarity.
	    In the isotropic case ($E=1$, center subfigures), the flow field is symmetric relatively to the swimmer axis.
	    Therefore, no reorientation occurs.
	    For $E>1$ (left subfigures), the flow is significantly more pronounced along the director axis (vertical).
	    This results in asymmetric flow fields relatively to the swimmer axis and effective shear flows around the swimmer body.
	    In $(a)$, for a pusher, this leads to a counterclockwise  and in $(d)$, for a puller, to a clockwise rotation of the swimmer.
	    Vice versa, for $E<1$ (right subfigures), the flow perpendicular to the director is more pronounced.
	    This implies opposite asymmetry and then opposite sense of rotation.
	   }
	   \label{quiverPlots}
\end{figure*}

For illustration of the resulting behavior, we present in Fig.~\ref{quiverPlots} quiver plots in addition to color contour diagrams of the self-induced velocity field given by Eq.~\eqref{selfInducedFlowFieldSwimmer} for a pusher [$f>0$, subfigures $(a)$, $(b)$, and $(c)$] and for a puller [$f<0$, subfigures $(d)$, $(e)$, and $(f)$] for three different values of the viscosity ratio~$E$.
The flow velocities are scaled by $f/(\nu_3 L)$. 
Here, the swimmer is inclined by an angle $\psi=\pi/6$ relatively to the horizontal.
We use the same parameters as in Fig.~\ref{SwimmerTraj}, where $\bar{\nu}=0$ and $\alpha=2/5$.

It becomes clear from Fig.~\ref{quiverPlots} that the viscosity ratio~$E$ has a pronounced influence on the resulting flow field induced by the swimmer. Moreover, we can illustratively understand from these plots the orientational behavior of the swimmer calculated above. 

For $E=1$, i.e., in the isotropic situation, corresponding to the center subplots $(b)$ and $(e)$ in Fig.~\ref{quiverPlots}, the induced fluid flow is symmetric with respect to the swimmer axis passing through both active force centers. Thus the swimmer propels in a straight way. 

However, the anisotropic environment for $E\neq1$ can break this symmetry. We noted above that $E>1$ is connected to a situation in which motion along the director is facilitated when compared to the transverse motion. Correspondingly, in subfigures $(a)$ and $(d)$ for $E>1$, the fluid flow induced by both active force centers is more pronounced along the director (i.e., here, along the vertical). For the pusher in subfigure~$(a)$, therefore, the flow induced by the force center on the right-hand side has a bias towards the top, whereas a bias towards the bottom arises around the force center on the left-hand side. In between, the overall fluid flow thus shows a shear component, which here contains a rotational component of counterclockwise sense. Consequently, the pusher in~$(a)$ with its body between the two force centers is rotated towards the director (i.e., here, towards the vertical). For the puller in subfigure~$(d)$, all flow directions are reversed, and therefore a clockwise rotation away from the director and towards a perpendicular orientation results. 

Vice versa, we motivated above that $E<1$ is connected to a situation in which motion perpendicular to the director is facilitated. Correspondingly, subfigures~$(c)$ and~$(f)$ for $E<1$ indicate a fluid flow asymmetric with respect to the swimmer axis and biased along the transverse axis (i.e., here, along the horizontal). More in detail, for the pusher in subfigure~$(c)$, the fluid flow induced by the force center on the right-hand side is more pronounced towards the right, whereas for the force center on the left-hand side it is more pronounced towards the left. Thus, in between, a net rotational component of clockwise sense arises, rotating the swimmer towards the axis perpendicular to the director. Again, for the puller in~$(f)$, all flows and thus rotational components are reversed so that the puller here rotates towards the director. 

% % % % % % % % % % % % % % % % % % % % % % % % lambda Effeckt

\subsection{Effect of~$\bar{\nu}$}

\begin{figure*}
\begin{center}
\includegraphics[scale=1]{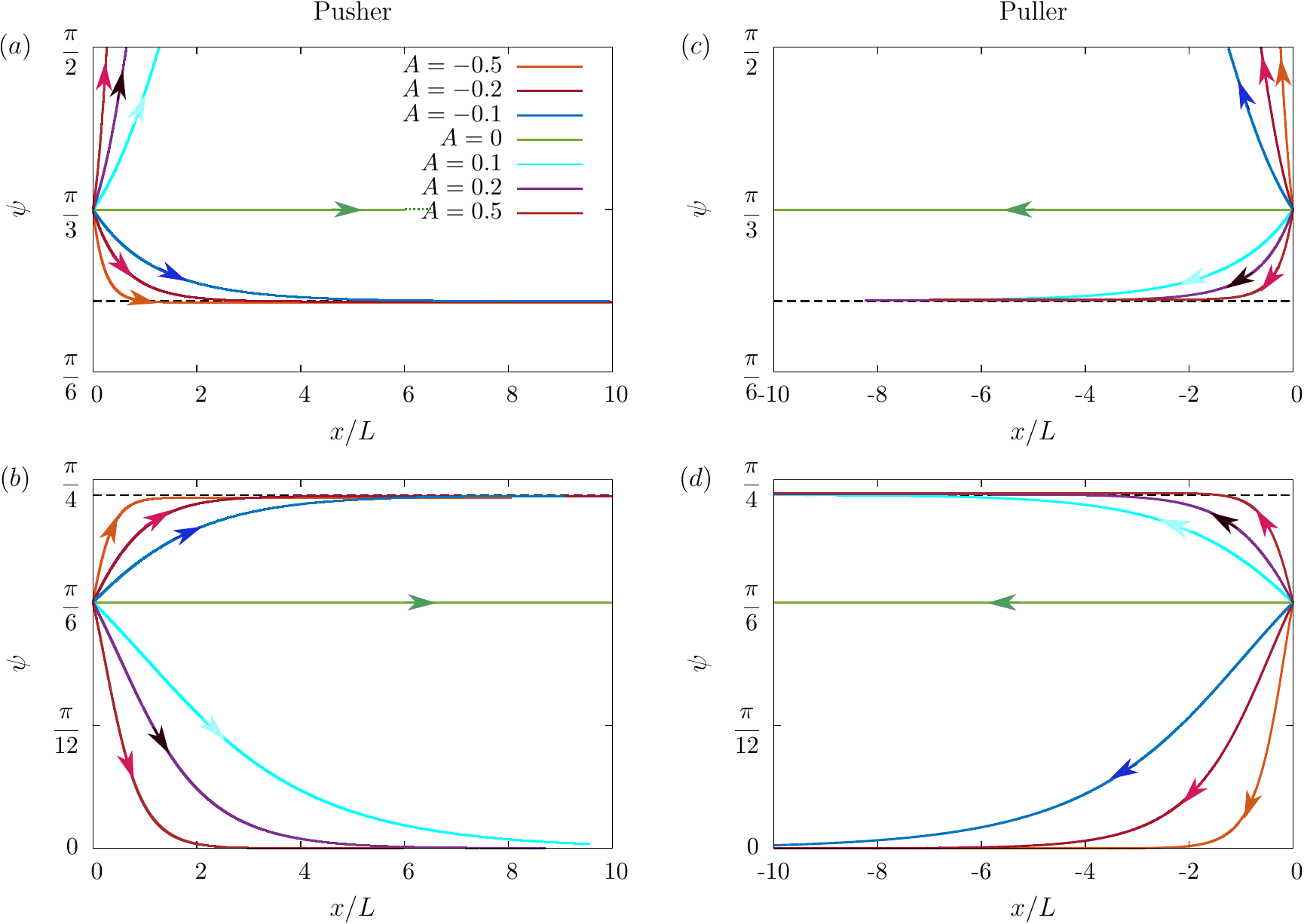}
\end{center}
\caption{(Color online) Variation of the inclination angle versus the scaled horizontal distance for a pusher- [subfigures $(a)$ and $(b)$] and a puller-type [subfigures $(c)$ and $(d)$] swimmer released from the origin with an initial inclination of $\psi_0=\pi/3 > \psi_A$ [subfigures $(a)$ and $(c)$] and $\psi_0=\pi/6 < \psi_A$ [subfigures $(b)$ and $(d)$] for $\varphi=0$ and various values of the viscosity ratio~$A$, while $E=1$.
Here, we set $\alpha=2/5$.
Depending on the propulsion mechanism and on the values of $A$ and $\psi_0$, the swimmer tends to align parallel or perpendicular to the director, or tends to swim along the axis given by $\psi_A \simeq 0.24\pi$ indicated by the dashed lines.
Arrows indicate the direction of time evolution.
When the steady alignment parallel to the director is reached $(\psi=\pi/2)$, the swimmer moves along the $z$-axis. This is why the corresponding curves end at a certain point in the $(x,\psi)$-plane.}
\label{SwimmerTrajAEffect}
\end{figure*}

We now consider a more general situation and allow for a small non-zero value of~$\bar{\nu}$.
Accordingly, we define the dimensionless number
\begin{equation}
	A = \frac{\bar{\nu}}{\nu_3} \, ,
\end{equation}
noting that $A \ge E-4$~\cite{forster71}.
Expanding the integrands in Eqs.~\eqref{greenFctsRealSpace_Main} perturbatively in the parameter~$A$ and evaluating the resulting integrals analytically, the solution for the Green's function corresponding to the plane~$y=0$ up to $\bigO \left(A^3\right)$ reads
%\begin{widetext}
\begin{subequations}\label{greensFct_smallA}
	\begin{align}
		\G_{zz} &= \frac{1}{8\pi\nu_3 R} 
		\bigg( 2-\sz^2 - \frac{A}{8} \, \sz^2 \left( 8-12\sz^2+5\sz^4 \right) %\notag \\
		+\frac{3A^2}{128} \, \sz^4 \left( 4-3\sz^2 \right) \left( 8-14\sz^2+7\sz^4 \right)	\bigg) \, , \\
		\G_{xx} &= \frac{1}{8\pi\nu_3 R} \bigg(  1+\sz^2
         + \frac{2}{\sz^2} 
         \left( \sqrt{\cz^2 + \frac{\sz^2}{E}}-1 \right) %\notag \\
         -\frac{5A}{8} \, \sz^4 \cz^2 
         +\frac{7A^2}{128} \, \sz^6 \cz^2\left(10-9\sz^2\right)
          \bigg) \, , \\
		\G_{xz} &= \frac{\cz \sz \cos\varphi}{8\pi\nu_3 R} 
		\bigg( 1-\frac{A}{8} \sz^2 \left(1+5\cz^2\right) %\notag \\
		+\frac{ \sz^4 A^2}{128} \left(80-140\sz^2+63\sz^4\right)
		\bigg) \, .
	\end{align}
\end{subequations}

Again, $\G_{zx} = \G_{xz}$.
{We remark that for commonly used nematic liquid crystals, the magnitude of $A$ is not necessarily small. Therefore, we have tested the direct numerical solution of Eqs.~\eqref{greenFctsRealSpace_Main} versus our analytical results in Eqs.~\eqref{greensFct_smallA}. Setting, for example, $E=1$, we find very good agreement for $|A|<1$ in the full range of $0\leq s^2 \leq 1$. 
For the commonly used liquid crystals MBBA and 5CB, we estimate a value of $A\approx4$ from the literature~\cite{stewart04}. 
Again testing the quality of Eqs.~\eqref{greensFct_smallA} within the full range of $0\leq s^2 \leq 1$, we obtain maximum deviations of $\mathcal{G}_{zz}$, $\mathcal{G}_{xx}$, and $\mathcal{G}_{xz}$ by factors of 1.3, 1.1, and 1.6 for $A=4$ and $E=1$, always with identical sign. 
Thus, even if the analytical expressions do not imply a full quantitative solution in several practical situations, they should still predict the correct qualitative behavior in a broad regime of accessible parameter values.}
As required, Eqs.~\eqref{greensFct_smallA} reduce to Eqs.~\eqref{greenFctsLambdaZero} for $A=0$.
Using Eqs.~\eqref{selfInducedFlowFieldSwimmer} and \eqref{defenitionVOmega}, the resulting swimming velocities and rotation rate are calculated up to second order in~$A$ as 
\begin{subequations}\label{swimmingVelocitiesGeneralized}
	\begin{align}
		V_x &= V_0 \cos\varphi\cos\psi \left(
	    \frac{w}{\cos^2\psi} -\tan^2\psi
		-\frac{3A}{8} \, \cos^2\psi \sin^2\psi
		+\frac{5A^2}{128} \, \cos^4\psi \sin^2\psi \left(8-7\cos^2\psi\right) \right)
		\, , \label{VxGeneralized} \\
		V_z &= V_0 \sin\psi \left(1
		-\frac{A}{8} \, \cos^2\psi \left(4-3\cos^2\psi\right)
		+\frac{A^2}{128} \, \cos^4\psi \left( 48-80\cos^2\psi+35\cos^4\psi \right) \right) 
		\, , \label{VzGeneralized} \\
		\Omega_y &= \Omega_0 \cos\varphi\sin\psi
		\left(
		\sec\psi \left( 1- \frac{1}{w} \right) 
		+ \frac{A}{8} \, \cos\psi \left(15\cos^2\psi-8\right) 
		+\frac{A^2}{128} \, \cos^3\psi 
		\left( 192-520\cos^2\psi+315\cos^4\psi \right)
		\right)
		 \, . \label{OmegaYGeneralized}
	\end{align}
\end{subequations}

%\end{widetext}

\begin{figure*}
\begin{center}
\includegraphics[scale=1]{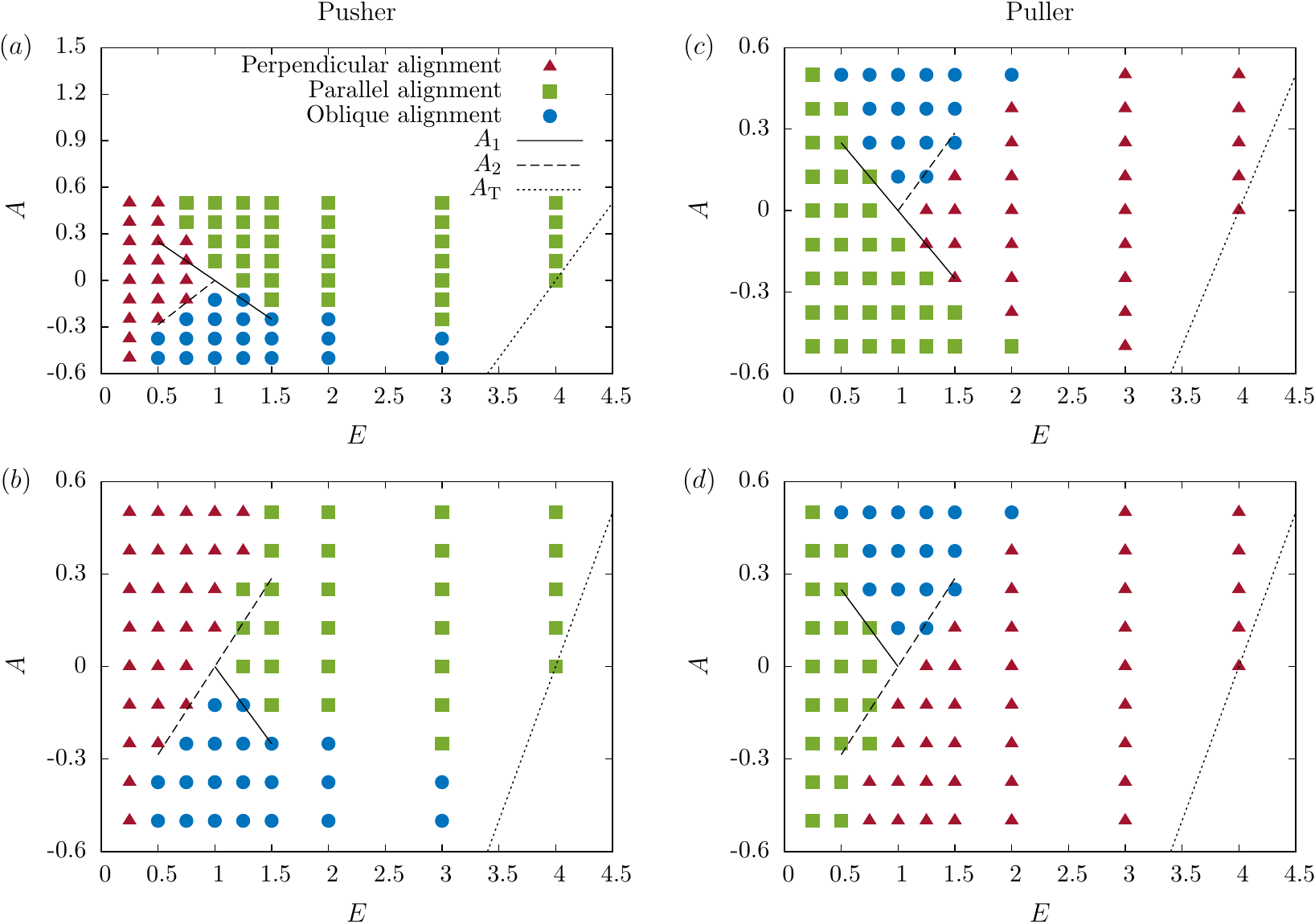}
\end{center}
\caption{(Color online) State diagram of orientational swimming behavior in a uniaxially anisotropic fluid for a pusher [subfigures $(a)$ and $(b)$] and a puller [subfigures $(c)$ and $(d)$] as obtained by numerically integrating the full nonlinear equations.
The swimmer is released from the origin with an initial inclination of $\psi_0=\pi/3$ [subfigures $(a)$ and $(c)$] and $\psi_0=\pi/6$ [subfigures $(b)$ and $(d)$] for $\varphi=0$.
The solid and dashed lines shown around the point $(A=0,E=1)$ are defined in Eq.~\eqref{definitionsA1UndA2}.
On the bottom right of each subfigure, the missing data points would fall into the unphysical regime of $A< A_\mathrm{T} = E-4$~\cite{forster71}.
}
\label{stateDiagramAE}
\end{figure*}

As before, to gain further insight, we first investigate the swimming behavior for a small deviation of the viscosity ratios $A$ and $E$ from the isotropic limit.
Correspondingly, linearizing the rotational velocity in Eq.~\eqref{OmegaYGeneralized} about $E=1$ and $A=0$ yields
\begin{equation}
	\Omega_y \sim \frac{\Omega_0}{4} 
	\left( 1-E-2A+\frac{15A}{4} \, \cos^2\psi \right) \sin \left(2\psi\right) 
	\cos\varphi\, . \label{Omega_y_general}
\end{equation}
The resulting dynamical system of equations has thus three trivial fixed points at $\psi=0$, $\pm \pi/2$, similarly as previously observed for the case $A=0$.
Defining
\begin{equation}
	\psi_A = \arccos \left( \frac{2 \sqrt{15 A \left(E-1+2A\right)}}{15|A|}  \right) \, , \label{psi_A}
\end{equation} 
two additional fixed points occur at $\psi = \pm \psi_A$ if $\psi_A$ is a real number.
Specifically, this is true for
\begin{eqnarray}
	A \ge \max\{A_1,A_2\} \quad \text{or} \quad A \le \min\{A_1,A_2\} \, , 
\end{eqnarray}
where we have defined
\begin{equation}
	A_1 =  \left(1-E\right)/2 \quad \text{and} \quad A_2 = 4\left(E-1\right)/7 \, .
	\label{definitionsA1UndA2}
\end{equation}

Determining the steady swimming trajectories for arbitrary values of the viscosity ratios~$E$ and $A$ is far from being trivial due to the existence of multiple fixed points depending on these values.
For convenience, we first consider the simple case of $E=1$, for which $\psi_A=\arccos \left(2\sqrt{30}/15\right) \simeq 0.24\pi$ is not a function of~$A$, and $A_1=A_2=0$. 
In addition, we confine ourselves for the sake of clarity to the case of $\psi_0 \in [0,\pi/2]$ and $\varphi=0$.
The other situations can then be deduced from symmetry arguments.

We now focus our attention on the behavior of a pusher-type swimmer, for which $\Omega_0>0$.
Considering an initial orientation $\psi_0>\psi_A$ in Eq.~\eqref{Omega_y_general}, then $\Omega_y < 0$ if $A>0$, and $\Omega_y >0$ if $A<0$.
Accordingly, the swimmer tends to rotate toward the director $(\psi = \pi/2)$ for $A>0$, while it tends to rotate toward the orientation set by $\psi_A$ for $A<0$.
In contrast to that, if $\psi_0 < \psi_A$, then the swimmer tends to rotate toward the axis perpendicular to the director $(\psi = 0)$ for $A>0$ and toward the axis given by~$\psi_A$ for $A<0$. 
An analogous discussion can be carried out for a puller-type swimmer, for which $\Omega_0<0$ and thus all these tendencies of rotation are reversed.

In Fig.~\ref{SwimmerTrajAEffect} we show the variation of the inclination angle~$\psi$ versus the scaled horizontal distance~$x/L$ for a microswimmer that is initially released from the origin in an anisotropic fluid for various values of $A$, while $E=1$.
Results for both a pusher [$(a)$ and $(b)$] and a puller [$(c)$ and $(d)$] are shown for two initial orientations, namely for $\psi_0=\pi/3 > \psi_A$ [$(a)$ and $(c)$] and for $\psi_0=\pi/6 < \psi_A$ [$(b)$ and $(d)$].
The curves were obtained by numerical integration of the equations of motion using the exact Green's function given by Eqs.~\eqref{greenFctsRealSpace_Main}.
We set the asymmetry parameter of the swimmer~$\alpha=2/5$.
Depending on the swimmer type, the values of the viscosity ratio~$A$, and the initial orientation, the swimmer aligns parallel to the director, perpendicular to the director, or maintains a steady orientation at a constant inclination angle~$\psi_A \simeq 0.24\pi$ that is independent of the value of~$A$.
The linearized analysis described above provides a useful framework for determining the steady-state swimming behavior.

Finally, we consider the general situation and vary both $E$ and $A$.
We begin with the situation of $A \ge \min \{A_1,A_2\}$ and $A \le \max \{A_1,A_2\}$, for which $\psi_A$ is not defined and thus the dynamical system has only three fixed points at $\psi=0$ and $\psi= \pm \pi/2$.
Moreover, let us consider a pusher-type swimmer such that $\Omega_0>0$, as well as initial orientations $\psi_0 \in [0,\pi/2]$ and $\varphi=0$.
For $E<1$, $\Omega_y >0$, while for $E>1$, it follows that $\Omega_y<0$ for all values of~$\psi$.
Therefore, for any initial orientation, a pusher will tend to rotate toward an axis perpendicular and parallel to the director, for $E<1$ and $E>1$, respectively.
A puller-type swimmer will exhibit the opposite behavior, as in the case of $A=0$ discussed earlier.

The situation of $A \ge \max \{A_1,A_2\}$ or $A \le \min \{A_1,A_2\}$, for which five fixed points occur, is more complex because the steady alignment also depends on the initial orientation.
If $A>0$ and $\psi_0>\psi_A$ or $A<0$ and $\psi_0<\psi_A$ [which is equivalent to $A \left(\psi_0-\psi_A\right) > 0$], again assuming initial orientations of $\psi_0 \in [0,\pi/2]$ and $\varphi=0$, we find from Eq.~\eqref{Omega_y_general} for a pusher $(\Omega_0>0)$ that $\Omega_y<0$.
Therefore, a pusher tends to rotate toward the director if $\psi_0>\psi_A$ and $A>0$.
It tends to rotate toward the axis set by~$\psi_A$ for $\psi_0<\psi_A$ and $A<0$.
This rotational sense is reversed when $A \left(\psi_0-\psi_A\right) < 0$, for which $\Omega_y>0$.
Then, the swimmer tends to rotate toward the axis set by~$\psi_A$ for $\psi_0>\psi_A$ and $A<0$, while it tends to rotate toward the axis perpendicular to the director if $\psi_0<\psi_A$ and $A>0$.
Naturally, again, switching to a puller reverses these tendencies.

A state diagram of swimming obtained by numerical integration of the governing equations using the exact Green's function given by Eq.~\eqref{greenFctsRealSpace_Main} is presented in Fig.~\ref{stateDiagramAE} for a pusher [subfigures $(a)$ and $(b)$] and a puller [subfigures $(c)$ and $(d)$] released from two exemplary initial inclinations $\psi_0=\pi/3$ [subfigures $(a)$ and $(c)$] and $\psi_0=\pi/6$ [subfigures $(b)$ and $(d)$].
The separating lines corresponding to the change in the number of fixed points obtained from the linearized rotational velocity are plotted according to Eq.~\eqref{definitionsA1UndA2} around the point corresponding to an isotropic fluid ($E=1$ and $A=0$). 
Far away from this point, it becomes necessary to account for the nonlinear contributions to the rotational velocity for an accurate determination of the lines of separation between the different states of orientational alignment.

{
Finally, on the basis of experimental data taken from the literature~\cite{stewart04}, we estimated the values $A \approx 4$ and $E \approx 1.7$ for the commonly employed nematic liquid crystals MBBA and 5CB.
Using these values, we find that a pusher-type swimmer tends to align parallel to the director for $\psi_0>\psi_A$ and perpendicular to the director for $\psi_0<\psi_A$, in agreement with the tendencies outlined in Figs.~\ref{SwimmerTrajAEffect} and \ref{stateDiagramAE}. 
Depending on the studied system in experiments, for instance, surface effects may favor the case of $\psi_0>\psi_A$~\cite{smalyukh08}, so that predominately parallel alignment is observed.
In contrast to that, we find a puller-type swimmer for both cases to tend to align along the oblique direction suggested by Fig.~\ref{SwimmerTrajAEffect}.}

\section{Conclusions}\label{sec:Schluss}

In summary, we have analyzed in this work theoretically the orientational behavior of self-propelled minimal model microswimmers in a uniaxially anisotropic fluid. Both pusher- and puller-type propulsion mechanisms were investigated. In general, we find different alignment behavior for these two types of swimmer. 

Our analysis started from the derivation of the Green's function describing fluid flows that are generated by point-like force centers in the anisotropic medium. To be able to perform the analytical calculations, we needed to {assume} that the anisotropy axis, i.e., the director when thinking of nematic liquid crystals, remains unperturbed and homogeneously aligned during the action of the swimmer. In a nematic liquid crystal, the latter may be achieved approximatively, for instance, deep in the nematic phase by strong external aligning fields. Then three remaining viscosities characterize the resulting fluid flows in the limit of fluid incompressibility. 

First, we did not consider the influence of the viscosity associated solely with elongational and compressional flows along the director. For the case corresponding to common nematic liquid crystals consisting, e.g., of rod-like molecules, in which the mobility of suspended particles is increased along the director when compared to perpendicular motion, we find an alignment of pusher-type microswimmers and their swimming paths with the director. This behavior is in agreement with recent experimental observations~\cite{kumar13,zhou14, sokolov15, genkin17, genkin18, mushenheim14b, mushenheim15}. For pullers, we obtain the opposite behavior, i.e., alignment and propulsion perpendicular to the director. Interestingly, our results predict a reversal of these alignment tendencies, both for pushers and for pullers, if we switch to the case of facilitated motion perpendicular to the director. Then the pushers orient themselves perpendicular and the pullers parallel to the director. We conjecture that this behavior could be observed, e.g., in discotic liquid crystals. Moreover, we showed how our results can be illustratively inferred from the fluid flows generated by the active microswimmers in the surrounding medium. 

Afterwards, we included the influence of the remaining third viscosity, i.e., the one connected only to elongational and compressional flows along the anisotropy axis. Interestingly, our results imply that in this case, besides orientations parallel and perpendicular to the director, also an oblique alignment relatively to the anisotropy axis becomes possible. Depending on the viscosity ratios and propulsion mechanism, this oblique alignment can be the one observed in the final steady state. Again, in each investigated situation, we found different orientational behavior for pushers and for pullers.

{At the end, let us recall that we have investigated in the present study the effects that the self-induced flows in an aligned anisotropic fluid can have on the orientational behavior of a self-propelling microswimmer. Depending on the system, for instance, concerning the shape of the swimmer, the type and magnitude of surface anchoring of the director on the swimmer body, or the nature of the anisotropic environment, elastic effects can become important as well~\cite{smalyukh08}. The relative magnitude of these contributions will be determined by the situation at hand.}

Altogether, in conclusion, our theoretical results and predictions further support the previously proposed objective of using anisotropic background fluids and nematic liquid crystals as host media to realize controlled and guided active transport. Since the director orientation in nematic liquid crystals can be imprinted in actual set-ups by suitable boundary conditions on confining container walls, requested paths of directed motion, also through complicated geometries, becomes conceivable. Moreover, the director orientation can be switched from outside by external fields, allowing for real-time control of the propulsion paths.

For the future, naturally, from a theoretical and analytical point of view, a severe challenge is to allow for deflections of the director field in our formalism and to include the corresponding couplings into the present description. {Additionally, the effect of higher moments of the force distribution exerted by the microswimmers on the surrounding fluid could be investigated. Particularly, the role of a force quadrupole for the alignment behavior should be analyzed.}
On the phenomenological and experimental side, interesting questions concern the possibility of observing the oblique states of alignment found in our work. Moreover, an increased attention to puller-type microswimmers in anisotropic fluids may be worthwhile. Apart from that, concerning our predictions for the varying viscosity ratios, experimental analysis of the behavior of active microswimmers in discotic liquid crystals offers a promising perspective.

\begin{acknowledgments}
We would like to thank Stephan Gekle, Achim Guckenberger, and the Biofluid Simulation and Modeling group at the University of Bayreuth where the boundary integral code extended in Appendix~\ref{appendix:BIM} has been developed for an isotropic fluid.
The authors gratefully acknowledge support from the DFG (Deutsche Forschungsgemeinschaft) through the projects DA~2107/1-1 (A.D.M.I) and ME~3571/2-2 (A.M.M.).
\end{acknowledgments}

% % % % % % % % % % % % % % % % % % % % % % % % % % % % % % % % % % % % % % % % % % % % % % % % % % % % % % % % % % % % % % % % % % % % % % % % % % % % % % % % % % % % % % % %

\appendix

\section{Inverse Fourier transforms}\label{appendix:inverseFourierTransformation}

The Fourier-transformed Green's function has been derived in Sec.~\ref{sec:GreenschenFunktionen} of the main body of the paper and is explicitly given by Eqs.~\eqref{GreenTensor_Fourier_ltz} and \eqref{GreenFunctions_Fourier_xyz}.
In this appendix, we provide technical details regarding the inverse Fourier transformation to yield the expressions for the Green's function in real space stated by Eqs.~\eqref{greenFctsRealSpace_Main}.

\subsection{The $zz$ component}

Substituting the $zz$ component of the Fourier-transformed Green's tensor as given by Eq.~\eqref{GreenTensor_Fourier_ltz} into Eq.~\eqref{FourierInverse} yields
\begin{equation}
	\G_{zz} = \frac{1}{(2\pi)^3} \int_{\mathbb{R}^3} \frac{k_\perp^2}{\nu_3 k^4+\bar{\nu} k_\parallel^2 k_\perp^2} \, e^{i \K \cdot  \vect{r}} \, \Intd \vect{k} \, . \label{mu_zz_def}
\end{equation}

To reduce the complexity of the integral, we make use of the change of variables $k_\parallel = k\cos\vartheta_k$ and $k_\perp = k \sin\vartheta_k$, where $\vartheta_k \in [0,\pi]$.
This leads to $\Intd \vect{k} = k^2\sin\vartheta_k \, \Intd k \, \Intd \varphi_k \, \Intd \vartheta_k$.
In addition, the argument of the exponential factor can be rewritten by noting that
\begin{equation}
 i \K \cdot  \vect{r} = ikR \big( \cos(\varphi_k-\varphi) \sin\vartheta \sin\vartheta_k + \cos\vartheta \cos\vartheta_k \big) \, , \notag 
\end{equation}
where $R := |\vect{R}| = |\R-\R_0|$, and $\varphi$ and $\vartheta$ stand for the azimuthal and polar angles, respectively, such that
\begin{equation}
	\vect{R} = \R-\R_0 
	= \left( 
	\begin{array}{c}
	x-x_0 \\
	y-y_0 \\
	z-z_0
	\end{array}
	 \right) 
	 = R \left( 
	 	\begin{array}{c}
	 	\sin\vartheta \cos\varphi \\
	 	 \sin\vartheta\sin\varphi\\
	 	\cos\vartheta
	 	\end{array}
	 	 \right)   \, .
\end{equation}

The first integration with respect to $\varphi_k$ between 0 and $2\pi$ can easily be performed by making use of the definition of the zeroth-order Bessel function \cite{abramowitz72},
\begin{equation}
   \frac{1}{2\pi} \int_0^{2\pi} e^{i\beta \cos (\varphi_k-\varphi)} \, \Intd \varphi_k = J_0(\beta) \, .
\end{equation}
The next integration with respect to $k$ between 0 and $\infty$ can then be performed by noting that
\begin{equation}
	\int_0^\infty e^{ibk} J_0(ak) \, \Intd k 
	= \begin{cases}
	\frac{1}{\sqrt{a^2-b^2}} & \text{~if~} |b|<|a| \, , \\
	\frac{i }{\sqrt{b^2-a^2}} \, \sgn (b) & \text{~if~} |a|<|b| \, , \\
	\end{cases}  \label{relation_Bessel0}
\end{equation}
wherein $\sgn$ denotes the sign function.
Using the change of variable $Q=\cos\vartheta_k/\sz$, Eq.~\eqref{mu_zz_def} can then conveniently be expressed as a convergent definite integral over $Q$ as
\begin{equation}
 \G_{zz} = \frac{1}{2\pi^2 R} 
 \int_{0}^{1} \frac{(1-\sz^2 Q^2) \, \Intd Q}{\big( \nu_3+\bar{\nu} \sz^2Q^2(1-\sz^2Q^2) \big) \sqrt{1-Q^2}} \, , \label{mu_zz_P}
\end{equation}
where we have abbreviated $\sz = \sin\vartheta$.
As already mentioned in the main text, an analytical evaluation of the resulting integral is possible only under certain conditions.
For $\bar{\nu}=0$, Eq.~\eqref{mu_zz_P} reduces to the isotropic result, namely
\begin{equation}
	\G_{zz} = \frac{1}{8\pi\nu_3 R} \left( 2-\sz^2 \right) \, .
\end{equation}

\subsection{The $xx$ and $yy$ components}

We next proceed in a similar way as done above for the $zz$ component to the $xx$ component of the Green's function.
Inserting Eq.~\eqref{G_xx} into Eq.~\eqref{FourierInverse} and making use of the identity
\begin{equation}
 \frac{1}{\pi} \int_0^{2\pi} \cos^2\varphi_k \, e^{i\beta\cos (\varphi_k-\varphi)} \, \Intd \varphi_k =  J_0(\beta)-J_2(\beta)\cos \left( 2\varphi \right)  
\end{equation}
together with Eq.~\eqref{relation_Bessel0}  and
\begin{equation}
	\int_0^\infty e^{ibk} J_2(ak) \, \Intd k 
	= \begin{cases}
	\frac{1}{a^2} \left( \frac{a^2-2b^2}{\sqrt{a^2-b^2}}+2ib \right) & \text{~if~} |b|<|a| \, , \\
	\frac{i \sgn (b)}{a^2} \left( 2|b| + \frac{a^2-2b^2}{\sqrt{b^2-a^2}} \right) &  \text{~if~} |b|>|a| \, ,
	\end{cases} \label{relation_Bessel2}
\end{equation}
we obtain
\begin{eqnarray}
	 \G_{xx} &=& \frac{1}{4\pi^2R} \int_0^{1} 
	     \bigg(\frac{\sz^2 Q^2 \Gamma_{-} }{ \nu_3+\bar{\nu} \sz^2 Q^2(1-\sz^2 Q^2)}  
	         + \frac{\Gamma_{+} }{ (\nu_3-\nu_2)\sz^2 Q^2+\nu_2 } \bigg)  \, \Intd Q \, ,
	  \label{mu_xx_P}
\end{eqnarray}
where, again, we have made the change of variable $Q=\cos\vartheta_k/\sz$.
Moreover, 
\begin{equation}
 \Gamma_{\pm} = \left( 1 \pm \frac{1-(2-\sz^2)Q^2}{1-\sz^2Q^2} \, \cos (2\varphi) \right) 
 \frac{1}{\sqrt{1-Q^2}} \, . \label{Gamma_MP}
\end{equation}
It is worth mentioning that the terms associated with $\Gamma_{-}$ and $\Gamma_{+}$ arise from the $ll$ and $tt$ related contributions to the Green's tensor, respectively.

The component $yy$ of the Green's function can readily be deduced from the $xx$ component by performing a quarter circle rotation of the frame of reference around the $z$-axis in the clockwise direction.
This corresponds mathematically to setting $\varphi \rightarrow \varphi-\pi/2$ in Eq.~\eqref{Gamma_MP}, which is equivalent to interchanging the meanings of $\Gamma_-$ and $\Gamma_+$ in Eq.~\eqref{mu_xx_P}.
Analytical expressions in the case $\bar{\nu}=0$ are possible but they are rather complex and lengthy and thus are omitted here.

\subsection{The off-diagonal components}

The $xz$ component of the Green's function can be obtained by inserting Eq.~\eqref{G_xz} into Eq.~\eqref{FourierInverse}, and integrating with respect to $\varphi_k$ and then $k$.
Making use of the identity
\begin{equation}
 \frac{1}{2\pi} \int_0^{2\pi} \cos\varphi_k \, e^{i\beta \cos(\varphi_k-\varphi)} \, \Intd \varphi_k = i\cos\varphi \, J_1(\beta) 
\end{equation}
together with
\begin{equation}
	\int_0^\infty e^{ibk} J_1(ak) \, \Intd k 
	= \begin{cases}
   \frac{1}{a} \left( 1+\frac{ib}{\sqrt{a^2-b^2}} \right) & \text{~if~} |b|<|a| \, , \\
   \frac{1}{a} \left( 1-\frac{|b|}{\sqrt{b^2-a^2}} \right) &  \text{~if~} |b|>|a| 
	\end{cases}
\end{equation}
for $a>0$, we find
\begin{equation}
 \G_{xz} = \frac{\cos\varphi}{2\pi^2 R} 
 \int_0^{1} \frac{\cz\sz Q^2 \, \Intd Q}{\big( \nu_3+\bar{\nu} \sz^2 Q^2(1-\sz^2 Q^2) \big) \sqrt{1-Q^2}} \, , \label{mu_xz_P}
\end{equation}
where $\cz= \cos\vartheta$.
In particular, we recover for $\bar{\nu}=0$ the component of the Green's function in an isotropic medium, namely
\begin{equation}
	\G_{xz} = \frac{\cz \sz \cos \varphi}{8\pi\nu_3 R} \, .
\end{equation}
The $yz$ component can readily be obtained by setting $\varphi \rightarrow \varphi-\pi/2$ in Eq.~\eqref{mu_xz_P}, or equivalently $\cos\varphi \to \sin\varphi$.

Finally, the $xy$ component of the Green's function can be obtained by inserting Eq.~\eqref{G_xy} into Eq.~\eqref{FourierInverse} to obtain
\begin{equation}
  \begin{split}
    \G_{xy} &= \frac{\sin (2\varphi) }{4\pi^2 R} 
    \int_0^{1} \bigg( \frac{1}{(\nu_3-\nu_2)\sz^2 Q^2+\nu_2} 
    -\frac{\sz^2 Q^2}{\nu_3+\bar{\nu} \sz^2 Q^2(1-\sz^2 Q^2)}  \bigg) 
              \frac{1-(2-\sz^2)Q^2}{(1-\sz^2 Q^2)\sqrt{1-Q^2} } \, \Intd Q \, ,
  \end{split}
\end{equation}
after making use of the identity
\begin{equation}
 \frac{1}{\pi} \int_0^{2\pi} \cos\varphi_k \sin\varphi_k \, e^{i\beta \cos(\varphi_k-\varphi)} \, \Intd \varphi_k = - \sin (2\varphi) \, J_2(\beta) \, ,
\end{equation}
in addition to Eq.~\eqref{relation_Bessel2}.
As already pointed out in the main body of the paper, $\G_{zx} = \G_{xz}$, $\G_{zy} = \G_{yz}$, and $\G_{yx} = \G_{xy}$ as required by the symmetry of the mobility tensor.

\section{Steady-state transverse distances covered by the swimmer until alignment is achieved}\label{appendix:deltaXZ}

In this Appendix, we calculate analytically the overall perpendicular and parallel distance covered until complete alignment parallel or perpendicular to the director is achieved, given an initial relative inclination~$\psi_0$ of the swimmer in the case of $\bar{\nu}=0$ studied in Sec.~\ref{subsection:OhneAlpha1}.

Posing $V_x = \Intd x/\Intd t$, $V_z = \Intd z/\Intd t$, and $\Omega_y=-\cos \varphi \, \Intd \psi/\Intd t$, the phase-space equations follow forthwith from Eqs.~\eqref{swimmingVelocities} upon eliminating the time differential.
Specifically,
\begin{subequations}
	\begin{align}
		\frac{\Intd x}{\Intd \psi} &=
		\frac{ \Lambda w \left( w - \sin^2\psi \right)}{\left( 1-w\right)\sin\psi} \, \cos\varphi \, , \label{diffXdiffPsi}
		 \\
		\frac{\Intd z}{\Intd \psi} &=
		\frac{\Lambda w \cos\psi}{1-w} \, , \label{diffZdiffPsi}
	\end{align}
\end{subequations}
where~$\Lambda$ has a dimension of length, defined as
\begin{equation}
	\Lambda =  \frac{2\alpha(1-\alpha)(1-2\alpha)L}{1-2\alpha+2\alpha^2} \, .
\end{equation}

For a given initial inclination~$\psi_0$, say, for simplicity, in the range $0 \le \psi_0 \le \pi/2$, the steady-state $x$-position of a swimmer that aligns parallel to the director can be determined.
Integrating Eq.~\eqref{diffXdiffPsi} for $\psi$ varying from $\psi_0$ to $\pi/2$ yields for the total distance $\delta_x$ traveled along the $x$-direction from the initial position to the final aligned state
\begin{equation}\label{delta_x}
		\begin{split}
			\delta_x &= \Lambda \cos\varphi \left( \cos\psi_0-\frac{\ln \left( \csc\psi_0-\cot\psi_0 \right)}{E-1}
			-\frac{\phi_- + \phi_+}{2 \sqrt{E} (E-1)} %\\
			%&
			+\frac{\arctan \left(\cfrac{\sqrt{q}}{w_0}\, \cos\psi_0 \right)}{E\sqrt{q}} \right) \, , 
		\end{split}
\end{equation}
wherein $q=1-1/E$ and $w_0 = w (\psi=\psi_0)$ according to Eq.~\eqref{defW}.
Moreover, we defined 
\begin{equation}
	\phi_\pm = \arctanh \left(  \cfrac{\sqrt{E}}{w_0} \left(q\cos\psi_0 \pm 1\right) \right) \, .
\end{equation}

The final $z$-position for a swimmer that aligns perpendicular to the director can be calculated in a similar way by integrating Eq.~\eqref{diffZdiffPsi} for $\psi$ varying from $\psi_0$ to~0.
We obtain for the total distance $\delta_z$ traveled along the $z$-direction from the initial position to the final steady state
\begin{equation}\label{delta_z}
	\begin{split}
		\delta_z &= \Lambda \left( \sin\psi_0 -\frac{1}{2q} \ln \left( \frac{2E \left(w_0 + \sin\psi_0\right)\sin\psi_0 + \cos^2\psi_0}{(1-\sin\psi_0)^2} \right) % \\
		%& 
		-\frac{1}{2 \sqrt{q}} \ln \bigg( E\Big( 1 - q \cos (2\psi_0)
		- 2w_0 \sqrt{q}\sin\psi_0 \Big) \bigg) \right) \, . 
	\end{split}
\end{equation}

Notably, both $\delta_x$ and $\delta_z$ diverge as $E\to 1$ because no alignment behavior occurs in an isotropic medium of equal viscosities $\nu_2=\nu_3$.
%\clearpage

% % % % % % % % % % % % % % % % %

\section{Hydrodynamic mobilities}\label{appendix:HydrodynamischeMobilitaeten}

The Green's function associated with a point force acting on the surrounding medium can be employed to assess the effect of a fluid on the dynamics of suspended particles, particularly for the computation of the self- and pair-mobilities~\cite{happel12, kim13}.
These are tensorial quantities that, for example, link the velocities of the particles to the forces exerted on them.
In the following, we derive explicit analytical expressions for the mobility functions in a uniaxial anisotropic fluid. 
They can, for instance, serve as a basis for future investigations of the behavior of some particle-based microswimmer models, such as the three-sphere swimmer introduced by Najafi and Golestanian and its different variations~\cite{najafi04, najafi05, golestanian08, golestanian08epje, golestanian09jpcm, ledesma12, pande17, liebchen18viscotaxis, daddi18, daddi18jpcm, lowen18}.
A comparison between analytical predictions and boundary integral simulations is also provided.

\subsection{Self-mobility function}

The hydrodynamic self-mobility function, denoted by $\boldsymbol{\mu}^\mathrm{S}$, relates the translational velocity~$\vect{V} $ of a particle to the force~$\vect{F} $ exerted on it.
Specifically,
\begin{equation}
	\vect{V} = \boldsymbol{\mu}^\mathrm{S} \cdot \vect{F} \, .
\end{equation}

The self mobility of a particle located at the origin is computed from the Green's function associated with the suspending medium as
\begin{equation}
	\mu_{ij}^\mathrm{S} = \frac{1}{(2\pi)^3} \int_{\mathbb{R}^3} \tG_{ij} (\K) \, \tilde{g}(\K) \, \Intd \K  \, , \label{mobilityDef}
\end{equation}
where $i,j \in \{x,y,z\}$ and $\tilde{g}(\K)$ is a wavenumber-dependent regularization kernel.
It is chosen in such a way as to consider only the wavenumbers $k:=|\K|$ that are smaller than the cutoff value $k_\mathrm{max} = \pi/(2a)$.
One way to set the regularization kernel is a Heaviside step function with a sharp Fourier cut-off of the form~\cite{levine01} 
\begin{equation}
  \tilde{g} (k) = H \left( \frac{\pi}{2a}-k \right) \, .
\end{equation}
However, this regularization kernel can cause a non-localized force distribution in real space in addition to the appearance of Gibbs oscillations in the radial velocity~\cite{gomez13}.
Therefore, a Gaussian regularization function of the form
\begin{equation}
	\tilde{g} (k) = e^{-(ka)^2/\pi} 
\end{equation}
is often preferred to overcome the drawback of the Heaviside regularization function.
Setting $k_\parallel = k\cos\vartheta_k$ and $k_\perp= k\sin\vartheta_k$ in the integrand, only $\tilde{g}(k)$ depends on the wavenumber~$k$, and thus both regularization kernels in our case lead to the same final result.

The obtained self-mobility tensor has the diagonal form
\begin{equation}
 \boldsymbol{\mu}^\mathrm{S} = \left(
		    \begin{array}{ccc}
		     \mu_\perp^\mathrm{S} & 0 & 0 \\
		     0 & \mu_\perp^\mathrm{S} & 0 \\
		     0 & 0 & \mu_\parallel^\mathrm{S}
		    \end{array}
		    \right) \, , \label{self-mobility-tensor}
\end{equation}
where we denote by $\mu_\perp^\mathrm{S}$ and $\mu_\parallel^\mathrm{S}$ the self-mobility function for the translational motion perpendicular and parallel to the director~$\vect{\hat{n}}$, respectively.
If we denote by $\vartheta_\mathrm{F}$ the angle between the applied force and the director, it follows from Eq.~\eqref{self-mobility-tensor} that the angle at which the velocity vector is directed relative to the director is
\begin{equation}
	\vartheta_\mathrm{V}
	= \arctan \left( \frac{\mu_\perp^\mathrm{S}}{\mu_\parallel^\mathrm{S}} \, \tan \vartheta_\mathrm{F} \right) \, .
\end{equation}
Clearly, the force and velocity vectors are collinear (so that $\vartheta_\mathrm{V}=\vartheta_\mathrm{F}$) in the particular situation of an isotropic fluid, for which $\mu_\perp^\mathrm{S} = \mu_\parallel^\mathrm{S}$.

\subsubsection{Parallel to the director}

The particle self-mobility function associated with the motion parallel to the director is readily obtained by inserting the component $\tG_{zz}$ of the Green's function from Eq.~\eqref{GreenTensor_Fourier_ltz} into Eq.~\eqref{mobilityDef}.
After the change of variable $q=\cos\vartheta_k$, the self mobility can conveniently be expressed in terms of a definite integral over~$q$ as
\begin{equation}
  \mu_\parallel^\mathrm{S} =  \frac{1}{4\pi a} \int_0^1 
  \frac{1-q^2}{\nu_3 + \bar{\nu} q^2 (1-q^2)} \, \Intd q \, ,
\end{equation}
which, upon integration, leads to the exact result defined for $\bar{\nu}>-4\nu_3$
\begin{equation}
  \begin{split}
   \mu_\parallel^\mathrm{S} = \frac{1}{4\pi a\nu_3 B} 
  \bigg(   \left(C_+ + \frac{1}{C_+} \right) \arctan C_+ %\\
  - \left(C_- + \frac{1}{C_-} \right) \arctan C_-  \bigg) \, .
  \end{split}\label{mu_parallel_final}
\end{equation}
Here, we set the scaled viscosity coefficients
\begin{equation}
  A = \frac{\bar{\nu}}{\nu_3}  \, , \quad
  B = \sqrt{A^2 + 4A } \, , \quad
  C_\pm = \sqrt{\frac{A \pm B}{2}} 
\end{equation}
and mention that $\arctan (iy) = i \arctanh y$, for $y \in \mathbb{R}$. 
Particularly, the Stokes mobility is recovered when $\bar{\nu}=0$,
\begin{equation}
	\left. \mu_\parallel^\mathrm{S} \right|_{\bar{\nu}=0} = \frac{1}{6\pi a \nu_3} \, . 
\end{equation}

Performing a Taylor expansion of Eq.~\eqref{mu_parallel_final} up to second order around the isotropic values corresponding to $\bar{\nu}=0$ and $\nu_3=\eta$ yields
\begin{eqnarray}
			6\pi\eta a \mu_\parallel^\mathrm{S}
			&=&
			 1-\frac{4}{35} \frac{\bar{\nu}}{\eta} -\left(\frac{\nu_3}{\eta}-1\right) 
			 +\frac{8}{385} \left(\frac{\bar{\nu}}{\eta}\right)^2 %  \notag  \\
			 %&&{}
			 + \left(\frac{\nu_3}{\eta}-1\right)^2 
			 +\frac{8}{35} \left(\frac{\nu_3}{\eta}-1\right) \frac{\bar{\nu}}{\eta}  \, . \label{mu_para_asympt}
\end{eqnarray}

\subsubsection{Perpendicular to the director}

The particle self-mobility function for the motion perpendicular to the director can be calculated by inserting the $xx$ component of the Green's function, given by Eq.~\eqref{G_xx}, into Eq.~\eqref{mobilityDef} to obtain
\begin{equation}
	 \mu_\perp^\mathrm{S} = \frac{1}{8\pi a}  \int_0^1 \bigg( \frac{1}{(\nu_3-\nu_2)q^2 + \nu_2} 
	             +  \frac{q^2}{\nu_3 + \bar{\nu} q^2 (1-q^2)} \bigg)  \Intd q \, , 
\end{equation}
where again the change of variable $q=\cos\vartheta_k$ has been made.
Integration yields the exact result
\begin{equation}
 \begin{split}
   \mu_\perp^\mathrm{S} = \frac{1}{8\pi a} 
  &\bigg( \frac{1}{\nu_2} \frac{\arctan \beta}{\beta} %\\
  %&
  + \frac{1}{\nu_3 B} \left( \frac{\arctan C_-}{C_-} - \frac{\arctan C_+}{C_+} \right) \bigg) \, , 
 \end{split}\label{mu_perp_final}
\end{equation}
where we have defined
\begin{equation}
  \beta = \sqrt{\frac{\nu_3}{\nu_2} - 1} \, .
\end{equation}
For $\bar{\nu}=0$, Eq.~\eqref{mu_perp_final} simplifies to
\begin{eqnarray}
	 \left. \mu_\perp^\mathrm{S} \right|_{\bar{\nu}=0} =
	 \frac{1}{8\pi a \nu_3} \left( \frac{1}{3} 
	 + \frac{\arctan \sqrt{\frac{1}{E}-1}}{\sqrt{E(1-E)}} \right) \, .
\end{eqnarray}
In the limit of equal viscosities, i.e., for $E=1$, we recover the particle bulk mobility in an isotropic fluid.

Performing a Taylor series expansion of Eq.~\eqref{mu_perp_final} up to second order about $\bar{\nu}=0$ and $\nu_2=\nu_3=\eta$ yields
\begin{eqnarray}
	6\pi\eta a \mu_\perp^\mathrm{S}
	&=& 1-\frac{3}{70}\frac{\bar{\nu}}{\eta}-\frac{1}{2}\left( \frac{\nu_2}{\eta}-1 \right)
	-\frac{1}{2}\left( \frac{\nu_3}{\eta}-1 \right) %\notag  \\
	%&&{}
	+\frac{2}{231}\left(\frac{\bar{\nu}}{\eta}\right)^2
	+\frac{2}{5} \left(\frac{\nu_2}{\eta}-1\right)^2
	+\frac{2}{5} \left(\frac{\nu_3}{\eta}-1\right)^2 \notag \\
	&&{}
	+\frac{3}{35}\frac{\bar{\nu}}{\eta} \left( \frac{\nu_3}{\eta}-1 \right)
	+\frac{1}{5} \left( \frac{\nu_2}{\eta}-1 \right) \left( \frac{\nu_3}{\eta}-1 \right) \, . %\notag
\end{eqnarray}
{Expressions of the self mobilities have likewise been obtained in Refs.~\onlinecite{gomez13, cordoba16}.}

\subsection{Pair-mobility function}

The fluid-mediated hydrodynamic interactions between particles are commonly expressed in terms of the pair-mobilities.
These are tensorial quantities that bridge between the velocity of one particle and the force exerted on another nearby particle.
Here, we restrict ourselves for simplicity to the translational pair-mobility function in the point-particle approximation.
The latter represents the leading-order term in an expansion of the pair mobilities in a power series of the ratio between particle radius~$a$ and the interparticle distance~$h$~\cite{swan07, swan10, aponte16}.
The accuracy and appropriateness of the point-particle approximation will be assessed hereafter by direct comparison with fully-resolved boundary integral simulations.

We now consider a pair of particles of identical radius~$a$, located at positions $\R_\alpha$ and $\R_\beta$, such that $h=|\R_\alpha - \R_\beta| \gg a$.
The induced translational velocity of particle $\alpha$ due to a force exerted on particle $\beta$ is approximated in terms of the pair-mobility tensor as
\begin{equation}
	\vect{V}_\alpha = \boldsymbol{\mu}^{\mathrm{P}}_{\alpha\beta} \cdot \vect{F}_\beta \, , 
	\qquad
	\boldsymbol{\mu}^{\mathrm{P}}_{\alpha\beta} = \boldsymbol{\G}\left( \R_\alpha-\R_\beta \right)
	+ \mathcal{O} \left( \epsilon^2 \right) \, , % \notag
\end{equation}
where $\epsilon = a / | \R_\alpha - \R_\beta | \ll 1$.

% % % % % % % % % % % % % % % % % % % % % % % % % % %

In the following, we calculate, for illustrative purposes, the components of the pair-mobility function for the two situations of the connecting vector $\vect{h} = \R_\alpha-\R_\beta$ between two particles aligned either parallel or perpendicular to the director.

First, we assume that particle $\beta$, upon which a force is exerted, is positioned at the origin, while particle $\alpha$ is located at $\vect{r}=(0,0,h)$.
In this configuration, the pair-mobility function for the motion parallel to the line of centers and $\vect{\hat{n}}$ can readily be obtained from Eq.~\eqref{mu_zz_P} as
\begin{equation}
    {\mu_{\parallel}}_{zz}^{\mathrm{P}} = \frac{1}{4\pi h \nu_3} \, .
    \label{mu_P_para_zz}
\end{equation}
Thus, we recover for $\nu_3=\eta$ the pair-mobility function in an isotropic fluid.

Next, we consider that particle $\alpha$ is located at $\vect{r}_\alpha = (\pm h,0,0)$ such that the line connecting the two centers is perpendicular to the director, still keeping the force on particle~$\beta$ along the director.
Setting $\vartheta=\pi/2$ yields
\begin{equation}
 {\mu_{\perp}}_{zz}^\mathrm{P} = \frac{1}{2\pi^2 h} \int_{0}^1 \frac{\sqrt{1-q^2}}{\nu_3 + \bar{\nu} q^2(1-q^2)} \, \Intd q \, .
 \label{mu_P_perp_zz}
\end{equation}
For an isotropic fluid, we have $\bar{\nu}=0$ as well as $\nu_3=\eta$, and thus ${\mu_{\perp}}_{zz}^\mathrm{P} = 1/(8\pi h \eta)$.
The latter is found to be half of the pair mobility parallel to the line of centers.
Physically, this expresses that it is much easier to move the fluid transversally than to squeeze it into or out of the gap separating the two particles~\cite{dufresne00}.

\begin{figure}
\begin{center}
\includegraphics[scale=1]{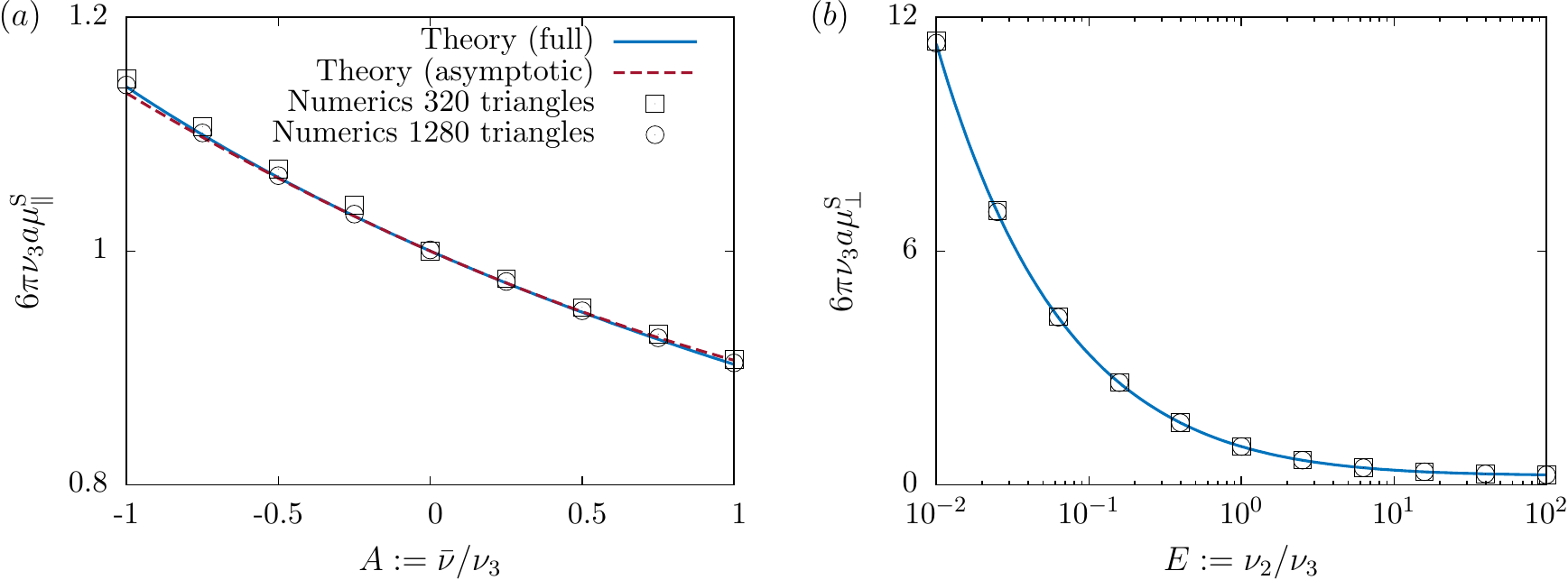}
\end{center}
\caption{(Color online) Variation of the rescaled self-mobility function for the motion $(a)$ parallel and $(b)$ perpendicular to the director versus the viscosity ratios.
Solid lines are the theoretical predictions given by Eqs.~\eqref{mu_parallel_final} and \eqref{mu_perp_final}, respectively, while the dashed line displayed in~$(a)$ is the asymptotic result given by Eq.~\eqref{mu_para_asympt} for $\eta=\nu_3$.
Symbols give the boundary integral simulation results for two different mesh resolutions.
Here we set $A=1/2$ in subfigure~$(b)$.}
\label{Self-Nematic}
\end{figure}

% % % % % % % % % % % % % %

The $xx$ components of the pair-mobility function, corresponding to the force on particle~$\beta$ and the resulting velocity of particle~$\alpha$ perpendicular to the director, for these two typical configurations can be obtained in an analogous way from Eq.~\eqref{mu_xx_P}.
We first consider the connecting vector~$\vect{h}$ of the two particles perpendicular to $\vect{\hat{n}}$ and then motion parallel to $\vect{\hat{n}}$.
Setting $\vartheta=\pi/2$ and $\varphi=0,\pi$, the term with the factor $\Gamma_{-}$ vanishes.
Accordingly, the pair mobility is solely determined from the $tt$ contribution to the Green's function, such that the $ll$-related part amounts to zero.
In this way, the pair mobility parallel~$(\parallel)$ to the line of centers can be evaluated analytically as
\begin{equation}
 {\mu_{\parallel}}_{xx}^{\mathrm{P}} = \frac{1}{4\pi h \sqrt{\nu_2 \nu_3}} \, .
 \label{mu_P_para_xx}
\end{equation}
Finally, by setting $\vartheta=0,\pi$ and $\varphi=0,\pi$, the pair mobility perpendicular~$(\perp)$ to the line of centers, with~$\vect{h} $ parallel to the director, reads
\begin{equation}
     {\mu_{\perp}}_{xx}^{\mathrm{P}} = \frac{1}{8\pi h \nu_2} \, .
     \label{mu_P_perp_xx}
\end{equation}
Again, the latter result is determined solely by the $tt$-related part in the Green's function.

% % % % % % % % % % % % % % % % % % % % %

\subsection{Comparison with numerical simulations}\label{appendix:BIM}

\begin{figure}
\centering
	\includegraphics[scale=1]{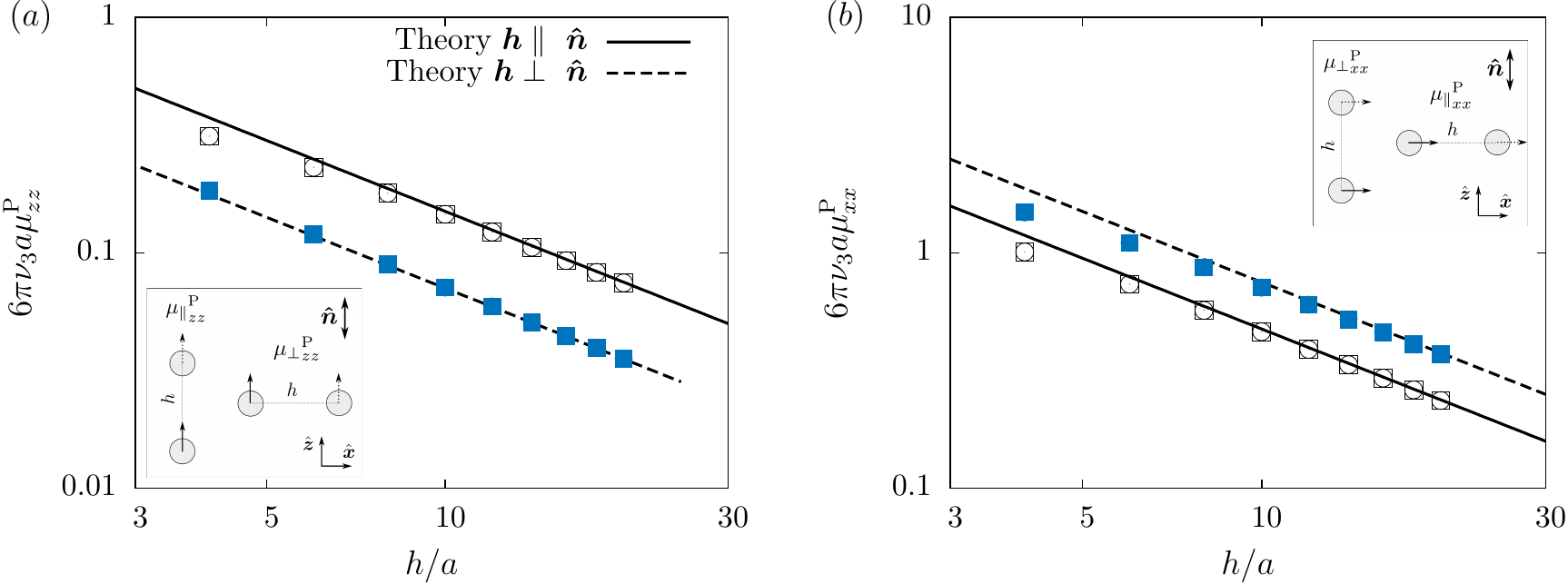}
\caption{Variation of the rescaled pair-mobility function for two particles aligned with their connecting vector~$\vect{h} $ parallel or perpendicular to the director $\vect{\hat{n}} \parallel \vect{\hat{z}}$ for $(a)$ the $zz$ and $(b)$ the $xx$ components.
Lines are the analytical predictions and symbols are the boundary integral simulation results obtained using 320 triangles (squares) and 1280 triangles (circles).
Here we set $E=1/10$ and $A=1/2$.
Insets: Illustration of typical configurations of a pair of particles disposed parallel or perpendicular to the director.  
The solid arrows indicate the force applied to the particle located at the origin, while the dotted arrows indicate the velocity of the particle located at distance $h$. }
\label{Pair-Nematic}
\end{figure}

In order to confirm our theoretical predictions and assess the range of validity of the expansions, we compare our analytical results with fully-resolved boundary integral method (BIM) simulations. 
The core idea of this numerical method is to expresses the solution of the governing equation given by Eq.~\eqref{Stokes} in terms of singularity distributions on the domain boundary~\cite{pozrikidis01}.
Then, the fluid flow field inside a control volume can be computed while requiring only knowledge of the traction on the domain boundaries.
The method has the special advantage that only a single 2D grid is required for the determination of the surface velocities at the boundaries as well as for the 3D computation of the flow field.
The BIM code used in this work was developed by the first author together with colleagues at the Biofluid Simulation and Modeling Group at the University of Bayreuth and has been validated in many flow problems in the Stokes regime~\cite{daddi16b, daddi18jcp, guckenberger16, Guckenberger_jpcm, daddi17d, daddi-thesis}.

Fig.~\ref{Self-Nematic} shows the rescaled self mobilities for the motion $(a)$ parallel and $(b)$ perpendicular to the director as functions of the viscosity ratios.
In the numerical simulations, the spherical particle is meshed by consecutively refining an icosahedron~\cite{krueger11, krueger12} for different triangulations.
Results for a coarse mesh with 320 triangles and a finer mesh with 1280 triangles are reported.
The asymptotic result given by Eq.~\eqref{mu_para_asympt} [shown as dashed lines in Fig.~\ref{Self-Nematic}~$(a)$] leads to a good estimate of the self mobility parallel to the director in the depicted range of~$A$.
As for the self mobility perpendicular to the director, shown for $A=1/2$ in Fig.~\ref{Self-Nematic}~$(b)$, both meshes brought about similar results.
The analytical predictions for both self mobilities are favorably compared with numerical simulations over the whole range of the considered viscosity ratios.

In Fig.~\ref{Pair-Nematic}, we present the pair-mobility function versus the interparticle distance~$h$ when the line connecting the centers of the two particles is oriented parallel or perpendicular to the director.
For $h > 5a$, the leading-order terms of the pair mobilities given by Eqs.~\eqref{mu_P_para_zz} through \eqref{mu_P_perp_xx} lead to a very good prediction.
However, for interparticle distances comparable with the particle radius, it may become necessary to account for the higher order terms for an accurate quantitative prediction of the fluid-mediated hydrodynamic interactions between the particles.

%\bibliographystyle{}
%\bibliography{biblio}

\end{document}